\documentclass[12pt]{article}

\usepackage[default]{lato}
\usepackage[T1]{fontenc}

\usepackage{scicite}
\usepackage{times}
\usepackage{xcolor}
\usepackage{graphicx}

\usepackage[utf8]{inputenc}

\newenvironment{sciabstract}{%
\begin{quote} \bf}
{\end{quote}}

\title{Uncovering commercial activity in informal cities} 

\author
{Daniel Straulino$^{1,2,*,+}$, Juan C. Saldarriaga$^{3,*}$, \\ Jairo A. G\'{o}mez$^{3,4}$ Juan C. Duque$^{3}$, Neave O'Clery$^{1,2}$\\
\\
\normalsize{$^{1}$Centre for Advanced Spatial Analysis, University College London, UK}\\
\normalsize{$^{2}$Mathematical Institute, University of Oxford, UK}\\
\normalsize{$^{3}$RiSE-group, Department of Mathematical Sciences,}\\
\normalsize{Universidad EAFIT, Medell\'in, Colombia}\\
\normalsize{$^{4}$i2t Research Group, Department of Communication and Information Technologies,}
\\
\normalsize{Universidad Icesi, Cali, Colombia.}\\
\\
\\
\normalsize{$^*$These authors contributed equally to this work.}
\\
\normalsize{$^+$To whom correspondence should be addressed: daniel.straulino@ucl.ac.uk.}
}
\date{}

\begin{document} 
\baselineskip24pt
\maketitle 

\begin{sciabstract}
Knowledge of the spatial organisation of economic activity within a city is key to policy concerns. However, in developing cities with high levels of informality, this information is often unavailable. Recent progress in machine learning together with the availability of street imagery offers an affordable and easily automated solution. Here we propose an algorithm that can detect what we call \textit{visible firms} using street view imagery. Using Medell\'in, Colombia as a case study, we illustrate how this approach can be used to uncover previously unseen economic activity. Applying spatial analysis to our dataset we detect a polycentric structure with five distinct clusters located in both the established centre and peripheral areas. Comparing the density of visible and registered firms, we find that informal activity concentrates in poor but densely populated areas. Our findings highlight the large gap between what is captured in official data and the reality on the ground.
\end{sciabstract}


\section*{Introduction}

The world is becoming more urban every day. According to the UN, 55\% of the world’s population is concentrated in  urban areas, a figure that is likely to increase to 68\% in the next thirty years \cite{un2018}. This process presents enormous social, economic and environmental challenges, particularly in the Global South, where most of the growth in urban population will take place. Despite high costs arising from congestion and density, it is widely accepted that the success of cities arises from dense social networks and access to a diversity of opportunity \cite{rosenthal2004evidence, duranton2005testing}. Most obviously this relates to access to jobs. 

The spatial organisation of jobs and economic activity \emph{within} developing cities is a core topic of concern across a large number of policy areas. Most straightforwardly, hubs of economic activity require services and transport connections to thrive \cite{bryan2020cities}. These connections facilitate, amongst other things, better labour market matching between workers and firms, and increase the accessibility of services and products to a wider catchment area \cite{goswami2019jobs}. Knowledge about job and firm density is also critically important in other domains such as urban planning and disaster resilience. However, in many developing city contexts, dominated by enormous informal sectors, data on the number, type or location of jobs and opportunities is either scarce or incomplete. 

According to the International Labour Organization, more than 60\% of the world’s working age population works informally. Most informal workers are found in developing countries where the proportion of informal workers is even higher, in some cases surpassing 90\% \cite{ilo2018}. Despite the size of the informal sector, the difficulty in tracking and reaching informal firms and workers means that there exists relatively little data on informal businesses outside of survey data (e.g., \cite{worldbank}) which is typically very expensive to collect, tends to be incomplete and suffer from reporting biases \cite{henley2009defining}. 

In response to these critical gaps, a burgeoning literature has emerging that aims to infer - both formal and informal - economic data from alternative sources. A well-known example is that of night lights, which has been used to estimate income growth \cite{henderson2012measuring}. Other approaches include crowd-sourced data \cite{crooks2015crowdsourcing, boeing2021spatial} and data on building height \cite{henderson2020building} to reveal urban form, and mobile-phone-derived commuting data to predict income \cite{kreindler2021measuring}. A large number of recent studies rely on imagery data which has recently become widely available, e.g., Landsat imagery, Google Maps, Kartaview, and Google Street View. Amongst many applications, satellite data has been deployed to predict consumption expenditure and wealth \cite{jean2016combining}, estimate enterprise counts \cite{goldblatt2019can}, monitor the growth of informal settlements \cite{kuffer2016slums}, and intra-urban poverty \cite{duque2015measuring}. Street imaging has been used to estimate the demographic makeup of neighbourhoods \cite{gebru2017using} and to classify buildings \cite{naik2017computer, gonzalez2020automatic, rueda2021use}. 

Here we propose a methodology to automatically detect and geo-reference the presence of commercial activities in a city. By using deep learning, we show that it is possible to efficiently identify what we call \textit{visible firms} from Google street imagery. These are firms that are easily identifiable as such at street level and include personal services, retail and amenities (bars, restaurants, etc.), and is sometimes referred to as 'street commerce' \cite{sevtsuk2020street}. Some sectors, which are less visible from the street, such as manufacturing, are not captured by our approach. Specifically, we use a manually labelled dataset of over 2000 panoramic images sourced from the metropolitan area of Medell\'in, Colombia to train a Neural Network to produce a dataset including the locations of over 170,000 visible firms across the city. It is important to mention that this methodology does not rely on detecting signs, which might not be present for certain businesses, but on the overall appearance of the facade, which includes exposed merchandise, architectural features and other signifiers of commercial activity. Furthermore, it does not distinguish between formal and informal firms, capturing both at once.   

While data on the spatial location of firms can be deployed for a very wide range of uses, we illustrate here how it can be used to uncover the patterns of concentration of economic activities in cities. This topic has been studied since the 1820s when cities were thought to consist of a dense Central Business District (CBD) surrounded by rings of progressively cheaper land \cite{berry1993challenges, gordon1986distribution}. More recently, this monocentric view has given way to a polycentric model \cite{giuliano1991subcenters, gordon1996beyond, lucas2002internal} consisting of multiple urban economic cores arising from increased mobility, suburbanisation and the movement of manufacturing to the periphery. Polycentrism is particularly acute in cities with weak internal transportation links \cite{ahlfeldt2015economics}. A number of studies suggest that global cities have become more polycentric over time \cite{mills1980comparison, bertaud2003spatial}, while evidence from the US suggests that polycentric metropolitan areas are larger and more dense \cite{arribas2014validity}. 

The forces behind firm clustering and industrial agglomeration in cities have been long studied. Benefits include easy access to customers and suppliers, shared labour supply and benefits from knowledge spillovers \cite{marshall1920economics, krugman1991increasing, rosenthal2004evidence, melo2009meta}. Costs include the cost of land and wages, which are likely to be higher in dense cities \cite{rosenthal2004evidence}. Distinct agglomeration patterns have been observed for manufacturing and services industries, with the latter benefiting from lower land costs and greater returns on access to labour and knowledge spillovers \cite{kolko2010urbanization, diodato2018industries}. Less is known about agglomeration forces, however, at a finer \emph{within city} spatial scale \cite{goswami2019jobs}. A small but growing number of studies suggest that agglomerative forces such as knowledge spillovers and inter-firm learning - key to the success of service firms - decline heavily with distance within cities (as reviewed by \cite{andersson2019economic}), thus providing a clear rationale for further research on the topic \cite{rosenthal2003geography}. 

A related literature focuses on the role of amenities in attracting people to cities \cite{bartik1987urban, albouy2016cities, sevtsuk2020street}. The presence of retail amenities is strongly dependent on the size of local population \cite{berry1988market}, while the share of specialised amenities correlates with city size \cite{glaeser2001consumer}. The 'consumer city' view sees the presence of amenities as drivers of growth and wages alongside traditional agglomeration economies \cite{glaeser2001consumer}. Central Place Theory provides a framework to understand the spatial organisation of amenities in cities from a consumer perspective \cite{christaller1966central}. The theory posits that amenities are organised in a manner in which multiple 'urban centers' serve consumers in the surrounding catchment area. A hierarchy of centers emerges as higher value goods and services, which are needed less frequently and for which customers will travel, concentrate in fewer clusters with large catchment areas. A large body of subsequent literature has adapted this approach, often adopting a less hierarchical polycentric-type model, e.g., \cite{meijers2007central}. Evidence from China suggests that polycentric cities enjoy more numerous and diverse amenities \cite{wang2020polycentric}.

Clustering and spatial agglomeration patterns have been less studied in developing city contexts. In particular, little is known about how agglomeration forces shape the locational decisions of informal firms \cite{overman2005cities, moreno2012critical, glaeser2017urban}. While the prevalence of polycentric cities is a well-established feature of developed cities, some studies have suggested that this model is also well-suited to Latin American cities for which formal employment is often distributed between a dense Central Business District and a few, relatively close, additional employment clusters (see \cite{krygsman2016cape} for a review). A number of studies find benefits for informal firms (productivity) and workers (wages) from density \cite{garcia2019agglomeration, tanaka2020agglomeration}. Investigating drivers of agglomeration, a small number of recent studies have shed some light on linkages between formal and informal firms, thought to hold the key to agglomeration economies in this context \cite{duranton2009cities}. These studies highlight the role of buyer-seller linkages and firm networks \cite{giuliani2005cluster, gebreeyesus2013innovation, mukim2015coagglomeration}. A significant result of these linkages and networks is technological/knowledge spillovers leading to learning and innovation for informal firms, as reviewed by \cite{giuliani2005cluster}.

In this paper, we focus on Colombia, where 76\% of its 47.1 million inhabitants reside in urban areas. A majority of Colombia's economy can be classed as informal. For example, estimates put the number of unregistered firms in Colombia above 50\% \cite{fernandez2018informalidad}. But there is wide variation across cities, with the average share of the working age population employed in formal employment standing at just 16.7\% \cite{o2018skill}. For the metropolitan area of Medell\'in, this share is significantly higher at 44\% (2015 data). Characterised by complex industry and international tourism alongside high levels of social segregation and informality \cite{lotero2016rich, o2018skill}, Medell\'in is a city of contrasts. In the last 25 years it has been subject to a dramatic urban transformation, particularly in the form of an innovative public transport system that includes cable cars to reach mountainous communities, that has attracted the interest of the international community \cite{arbaux2012Medellin}.

We focus on analysing the spatial pattern of visible firms detected in our dataset. We show that visible firms are spread throughout the metropolitan area. We apply LISA (spatial autocorrelation) analysis to the locations of visible firms and identify five distinct clusters of commercial activity located both in the traditional economic centre and in poorer areas on the periphery. This finding is consistent with the aforementioned, but limited, existing literature on the spatial distribution of economic activity - and particularly services and amenities - into polycentric cores. We complement our analysis by comparing our results with a data-set of formal firms registered with DIAN (the Colombian tax authority). Registered firms exhibit a significantly higher level of clustering, but are largely absent from much of the broader metropolitan area. We find just two formal clusters, one of which (the city centre) overlaps with one of the clusters found using the visible firm data. Taken together, our analysis challenges official data and previous work based on land values suggesting a monocentric structure in Medell\'in \cite{duque2013localizacion}. By comparing the density of visible firms to registered commercial firms, we infer the presence of informal firms and show that they concentrate around formal clusters, and in densely populated but poorer and less industrially complex parts of the city. Finally, we find that non-adherence by visible firms to commercial and mixed use land zoning is high across all socio-economic strata, and particularly high for visible firms located on the outskirts of formal clusters. 

\begin{figure*}[t!]
\includegraphics[width=\linewidth]{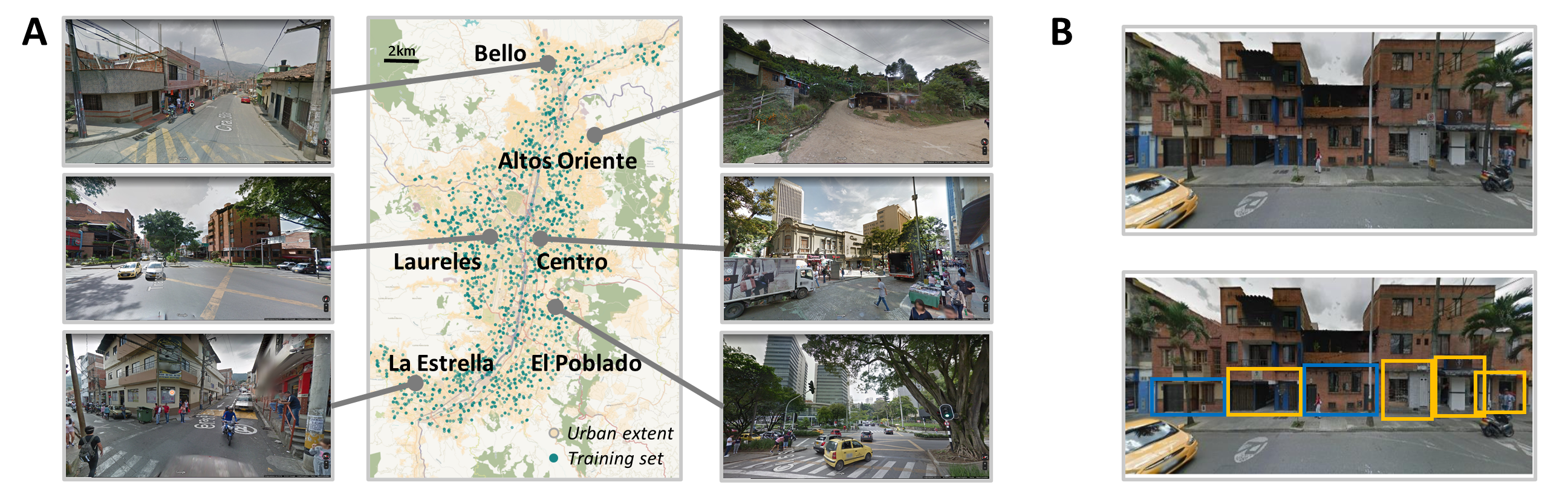}
\caption{\textbf{Street view imagery.} Commercial activity in Medell\'in comes in many different shapes and forms. \textbf{(A)} We illustrate this variety by showing shop fronts from six different locations in the urban area. To account for this diversity, it is important that we sample different types of facades. The map in the middle shows all the points in the city where we have sampled images to train our algorithm, which cover all regions of the metropolitan area. \textbf{(B)} For each point, a panoramic photography is obtained from Google Street View. These photographs are processed to obtain two images, one of each side of the road. Each image is then manually labelled: a bounding box is drawn for each facade, and is then marked as commercial (shown in yellow) or non-commercial (shown in blue).}
\label{commerce}
\end{figure*}

\section*{Results}

\subsection*{A new method to uncover visible commercial activity in data scarce contexts}

Here, we propose an algorithm based on machine learning applied to Google Street View images which enables us to identify the location of what we describe as visible firms, i.e., those that are easily identifiable from the street. For any region for which such images are available, this approach will produce a geo-referenced database of visible firms. Here, we apply the algorithm to the metropolitan area of Medell\'in, Colombia. A high level of informal economic activity and socio-economic diversity, as well as a rugged topography and a non homogeneous urban sprawl, combine to present a challenging case study for our detector. While our analysis focus on Medell\'in, the methodology is straightforwardly transferable to other cities and regions.

The workflow employed to build the algorithm is given in Figure \ref{FlowDetections}A, with a detailed overview given in Appendix \ref{detector}. We train the machine learning algorithm with a manually labelled dataset of over 8,000 facades corresponding to 2,000 randomly sampled locations in the region (illustrated in Figure \ref{commerce}). For each location we obtain a $360^{\circ}$ panoramic image which is transformed into two pictures, one for each side of the street. In each of this pictures a bounding box is drawn over each facade and labelled as either commerce or not (Figure \ref{commerce}B). We follow established procedures and augment the set of images \cite{shorten2019survey} before fitting a pre-trained Faster R-CNN \cite{ren2015faster}. 

The performance of the algorithm is summarized in Table 1. The precision score (i.e., the proportion of commercial facades that are correctly identified) is very high (>97\%) in both the validation and the test subsets, indicating that almost all the detections are visible firms. The recall (the proportion of visible firms that are detected) is lower, around 60\%, and thus some visible firms are likely to go undetected. This performance is comparable to related work \cite{yu2015large, ye2019urban, sharifi2020detecting} which obtain scores of around (85\%,65\%) for precision and recall. We also show the the F1 score which is the harmonic mean of the precision and recall scores.

\begin{center}
    \begin{tabular}{||c c c c||} 
        \hline
        Score & Training & Validation & Test \\ [0.5ex] 
        \hline\hline
        Precision & 0.9965 & 0.9706 & 0.9816 \\ 
        \hline
        Recall & 0.8871 & 0.6094 & 0.5941 \\
        \hline
        F1 & 0.9386 & 0.7487 & 0.7402 \\[0.1ex] 
        \hline
    \end{tabular}
    \begin{small} 
    \\
    \vspace{0.3cm}
    \textbf{Table 1.} Performance of the algorithm. \end{small}
\end{center}

\begin{figure*}[t!]
  \includegraphics[width=\linewidth]{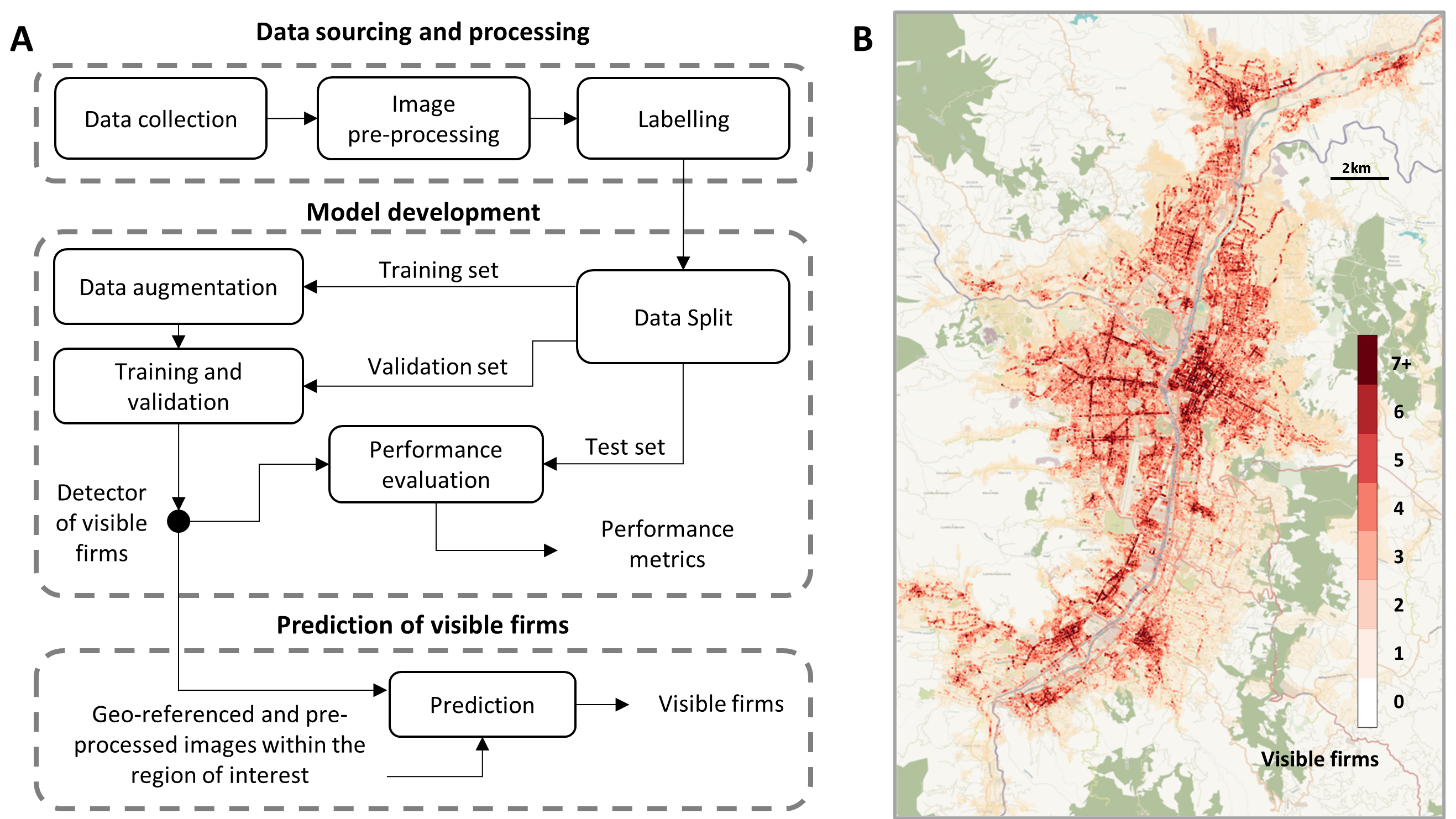}
  \caption{\textbf{Detecting visible firms.} \textbf{(A)} The workflow for detecting visible firms has three main stages. During data collection, the images necessary for training are acquired, processed and labelled. In the model development stage, the model is trained on an augmented dataset and the best hyper-parameters are found using a validation set, while the performance metrics are computed using a an (unseen) test set. Once the detector has been fitted, we apply it to the region of interest, in this case Medell\'in. \textbf{(B)} The set of visible firms in Medell\'in. The detection algorithm found over 170,000 visible firms in the metropolitan area; while they concentrate around the city centre and along some of the busiest streets, they are present across the whole urban sprawl.}
  \label{FlowDetections}
\end{figure*} 

We focus on the metropolitan area of Medell\'in as defined by the DANE (National Administrative Department of Statistics) in their most recent census (see Appendix \ref{data}), spanning 10 municipalities of the Antioqu\'ia department and home to 3.5 million people. In order to obtain a consistent picture of the number of visible firms, we exclusively used street view images captured in 2017. Figure \ref{FlowDetections}B shows the full set of 170,000 visible firms found by the detector superimposed on the metropolitan area of Medell\'in, shown in orange. As expected, concentrations of firms are clearly visible in this dataset. But we also notice that the footprint of the detections extends across most of the urban area. While the algorithm will miss a fraction of the true set of all visible firms (see recall score above), the overall spatial distribution is robust with respect to random omissions (see Appendix \ref{detector}). 

\begin{figure*}[t!]
  \includegraphics[width=\linewidth]{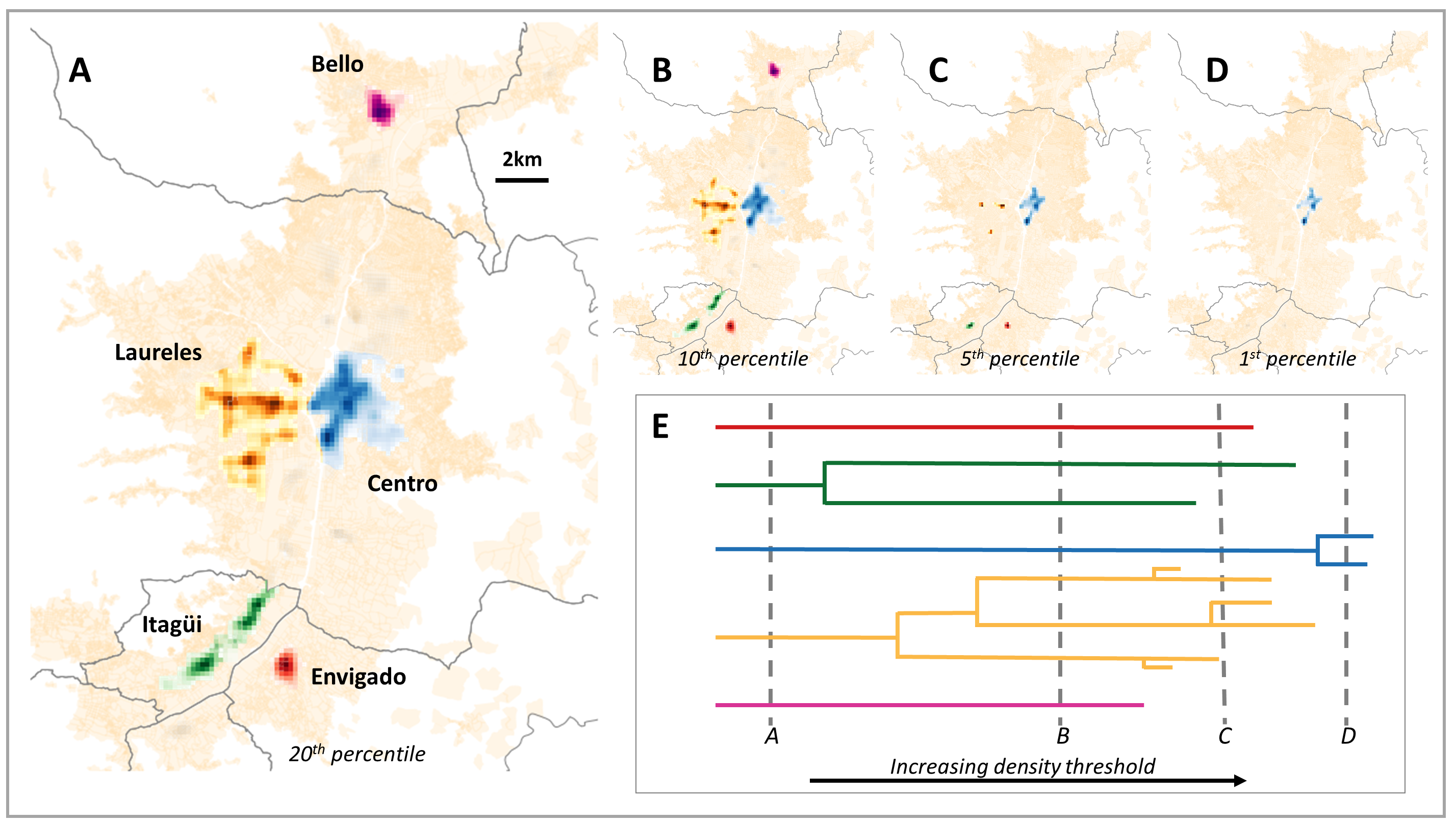}
  \caption{\textbf{Clustering of visible firms.} We apply LISA (local indicators of spatial association) statistics to identify clusters of visible firms across the metropolitan region. \textbf{(A)} When we include the top 20\% of cells by density value there are five distinct clusters (City center, Itagüi, Bello, Laureles, and Envigado). \textbf{(B-D)} By changing the density threshold at which we include cells in the analysis, we can distinguish clusters by the 'strength' of clustering and identify sub-divisions of clusters into smaller, more concentrated, agglomerations. \textbf{(E)} A dendrogram illustrates the emergence and merging of clusters as the density threshold changes. }
  \label{clusters}
\end{figure*}

\subsection*{Visible firms cluster around five centers in Medell\'in}

\begin{figure*}[t!]
  \includegraphics[width=\linewidth]{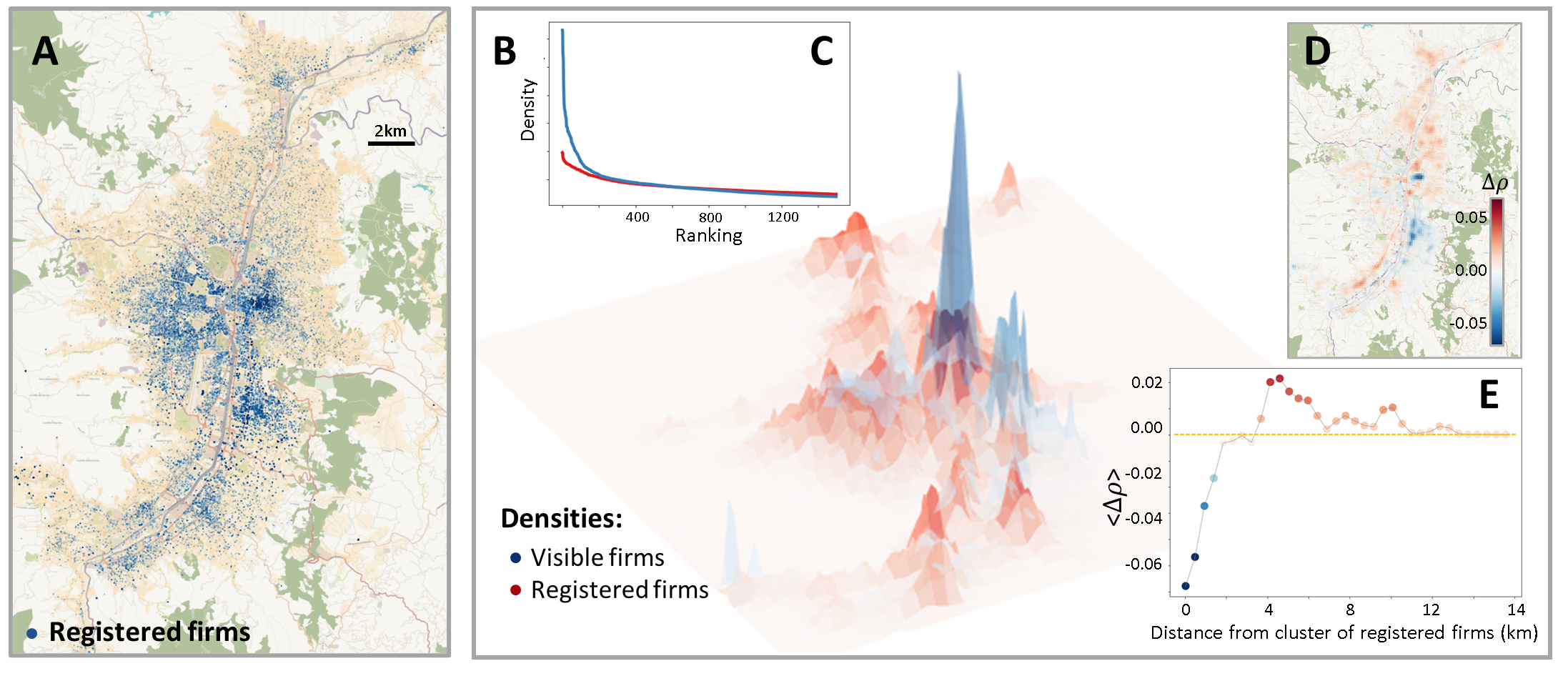}
  \caption{\textbf{Visible and registered firms.} \textbf{(A)} The location of registered firms in Medell\'in, where each firm is represented by a single blue dot. \textbf{(B)} The surface represents the spatial density of visible ($\rho_{vis}$ in red) and registered ($\rho_{reg}$ in blue) firms. The peaks of the blue distribution are much more pronounced, but the red distribution extends across the metropolitan area. \textbf{(C)} Ranking the cells according to density allows us to see how much more spatially concentrated registered firms are; the cells corresponding to the top percentile (1\%) accounts for over 20\% of the registered firms but only 13\% of the visible firms. \textbf{(D)} We calculate $\Delta\rho := \rho_{vis}-\rho_{reg_c}$, the surplus density of visible firms over registered commercial firms. Positive values of $\Delta\rho$ suggest regions with significant informal commercial activity. \textbf{(E)} Informal commercial activity peaks at a distance of 4km from the centroids of the formal clusters (El Poblado and the city centre).}
  \label{densities}
\end{figure*}

To investigate spatial clustering in the distribution of visible firms, we first estimate the density of visible firms across the metropolitan area. The density effectively smooths errors in the data, which mainly arise from distances between the point an image was taken and the precise location of a firm. Using a grid, we divide the region into cells of 200 m by 200 m (about the size of a city block), which determines the granularity of the density we will obtain. By applying kernel density estimation \cite{rosenblatt1956remarks} to the set of visible firms we obtain $\rho_{vis}$, which is shown as a red surface in Figure \ref{densities}B. This density can be interpreted as a spatial probability distribution for a firm sampled at random from our detections. 

Equipped with the density, we applied LISA (local indicators of spatial association) \cite{anselin1995local} statistics to identify clusters of visible firms across the metropolitan region (similar to \cite{arribas2014validity}). For each cell, LISA performs a statistical test of the spatial auto-correlation of $\rho_{vis}$. The resulting p-values can be used to decide which cells show a significant clustering pattern. A significance level of $p=0.1$ is commonly used as a threshold for identifying employment clusters \cite{arribas2014validity}. In Figure \ref{clusters} we adopt this p-value (with sensitivity test in Appendix \ref{clustersReg}), but vary the density value at which cells are included in the analysis.  

Figure \ref{clusters}A shows that when we include the top 20\% of cells (or higher, see Appendix \ref{clustersReg}) there are five distinct clusters (city center, Itagüi, Bello, Laureles, and Envigado). The city centre and Laureles are both located in the municipality of Medell\'in, while the other clusters are located in the old town centres of their respective municipalities. These clusters constitute evidence of the polycentric nature of Medell\'in in which several urban cores operate as economic engines within the city. 

We vary the density threshold to further investigate the spatial concentration of visible firms, in particular to distinguish clusters by the 'strength' of clustering and to identify sub-divisions of clusters into smaller, more concentrated, agglomerations. We can illustrate the evolving structure of clusters for different density thresholds via a dendrogram shown in Figure \ref{clusters}E. The visualisation reveals at what threshold value a cluster appears or merges with other clusters, thus uncovering sub-structures within clusters. 

We observe that as the threshold increases, Bello (a poor suburb in the north of Medell\'in) has the weakest clustering and is the first to disappear. In contrast, the Centro cluster is the strongest and persists as the threshold reduces. At very high values of the density threshold, only Centro remains. Laureles exhibits sub-structure as it fragments into several clusters (B-C) before disappearing (D). Similarly Itagüi splits into North and South, which remain present until Itagüi North disappears (C) followed by Itagüi South and Envigado (D).  

There is no doubt that the nature of the firms we capture tend to be service oriented, and that these tend to locate close to customers, particularly when cities are less well connected. Although recent investments in public transport have yielded a well connected metropolitan area, with travel times from Bello to Envigado cut from over 2 hours to just 30 minutes since the mid 1990s, it appears that economic activity remains highly distributed. This is likely in major part due to a path dependence in the development of economic clusters. In particular, notice that even though the area covered by the clusters increases as we increase the threshold, the clusters never cross municipality lines (shown in grey). Apart from Laureles, the clusters are located in the old centre of the municipalities. Hence, it appears that path dependence in the expansion of commercial activities has resulted in a modern configuration of economic clusters along historic municipal lines. 

Previous work on identifying commercial clusters in Medell\'in \cite{duque2013localizacion} used the land value of the locations of registered firms, aggregated to neighbourhood level, to produce clusters of industrial activity and services. The authors identified a single commercial cluster within the municipality of Medell\'in (they did not consider the wider metropolitan area), which contrasts with our analysis that uncovers two distinct clusters in this area, Laureles and Centro, which are separated by the river. Hence, our approach, which does not depend on official data and is not restricted by a particular geographical unit, produces more finely grained results.

\subsection*{Commercial informal (visible but non-registered) firms lie in the shadow of registered firms}

While previous work \cite{duque2013localizacion} has investigated the presence of economic clusters for the municipality of Medell\'in, this study omits the broader metropolitan area. Here we probe to what extent existing data on the location of formal registered firms captures the economic geography of Medell\'in. To do this we exploit the Colombian Statistical Directory of Companies (DEE, Directorio Estad\'istico de Empresas), a dataset that contains the location of all firms registered with DIAN (Direcci\'on de Impuestos y Aduanas Nacionales, the Colombian tax authority).

This dataset contains not only the location of each firm but also its industry code (at 4 digit level). The number of registered firms in the metropolitan area is around 150,000, which is smaller than the number of visible firms (170,000). Note, however, that this dataset only contains the locations of headquarters, and not branches, and so is under-counting the total number of registered establishments. Furthermore, the registered and visible firms datasets overlap in the sense that the visible firms dataset contains both commercial formal (registered) firms as well as commercial informal firms. Hence, a subset of firms - those that are both formal/registered and commercial - should appear in both datasets. Alongside the full set of registered firms, we use a list of industries labelled as 'street commerce' by \cite{sevtsuk2020street} to create a subset of registered firms (approx 15,000 firms) termed 'registered commercial firms'. These are active in industries that are likely to be visible from the street (see Appendix \ref{byindustry} for the list of industries). We note that this subset is an order of magnitude smaller than the visible firms dataset, further signalling the extent of missing commercial activity in official data. 

In Figure \ref{densities}A we show the distribution of registered firms (full set) in Medell\'in. We observe that their spatial distribution is different from that of the visible firms, most notably in the absence of registered firms in the northern poorer communities of Bello and Copacabana. We show the density of both the registered firms $\rho_{reg}$ and visible firms $\rho_{vis}$ (correlation of 0.64) in Figure \ref{densities}B. The high concentration of registered firms in the city centre is apparent from the large peak in the density. A second region of high density just south of the city centre is also apparent; it corresponds to the new business district (El Poblado) where sky-scrapers house the headquarters of many of the largest firms. By contrast, the density of visible firms is flatter across the urban extent. Hence, we capture a large swathe of economic activity outside the main centres that is not present in official data. 

To quantify the difference in concentration we rank the cells according to both $\rho_{vis}$ and $\rho_{reg}$. Figure \ref{densities}C shows that the the highest ranked cells account for a larger proportion of registered firms than they do for visible firms. For example, the cells representing the top percentile (1\%) account for 19.8\% of registered firms but only for 13.7\% of visible ones. We quantify the disparity in concentration by calculating the coefficient of variation (CV) for both densities. While the density of visible firms has a CV = 2.22, the density of registered firms scores 2.81, indicating a higher degree of concentration. If we restrict to commercial registered firms, we find an even higher concentration of 3.03.

Next, we apply LISA analysis as above to the set of registered firms. At the $p=0.1$ significance level, we find just two formal clusters, Centro and El Poblado (see Appendix \ref{regclust}). We repeat this exercise for the reduced set of 'registered commercial firms', finding again just these two clusters at $p=0.1$ significance level. Hence, when using official data on firms for Medell\'in, it appears that there exists just two centers of commercial activity (one of which overlaps with the visible firm clusters, Centro). This is in stark contrast to the five distinct centers that are apparent when LISA is applied to the visible firms dataset. Hence, we observe limited spatial overlap between concentrations of registered and visible firms outside the city centre which might suggest an absence of widespread linkages between the formal and informal sectors, as suggested by \cite{dominguez2019agglomeration}.

We further investigate the spatial concentration of informal activity relative to the formal clusters. While we cannot directly disentangle formal and informal firms in our dataset of visible firms, we can look for areas in which there is an 'excess' concentration of visible firms relative to registered commercial firms. To do this, we calculate the difference in density between visible and registered commercial firms $\Delta\rho:=\rho_{vis}-\rho_{reg_c}$ (Figure \ref{densities}D). Positive values (red) indicate that visible firms are more concentrated than registered commercial firms.

We compute the mean value of $\Delta\rho$ as a function of the distance to the formal clusters as shown in Figure \ref{densities}E. We observe a peak in the concentration of visible firms relative to registered commercial firms at around 4km from the centroid of the clusters. Hence, we find that unregistered or informal commercial firms concentrate in areas surrounding formal clusters. In the following section we will investigate the characteristics of these areas, including socio-economic status, population density and industrial complexity.

Overall, we find that visible firms are widely distributed across the urban extent and organised around multiple clusters distributed across the center, north and south of the city. By contrast, registered formal firms are concentrated in just two central clusters, Centro and El Poblado, and attract a surplus concentration of visible firms relative to registered commercial firms in their surrounding areas.

\begin{figure*}[t!]
  \includegraphics[width=\linewidth]{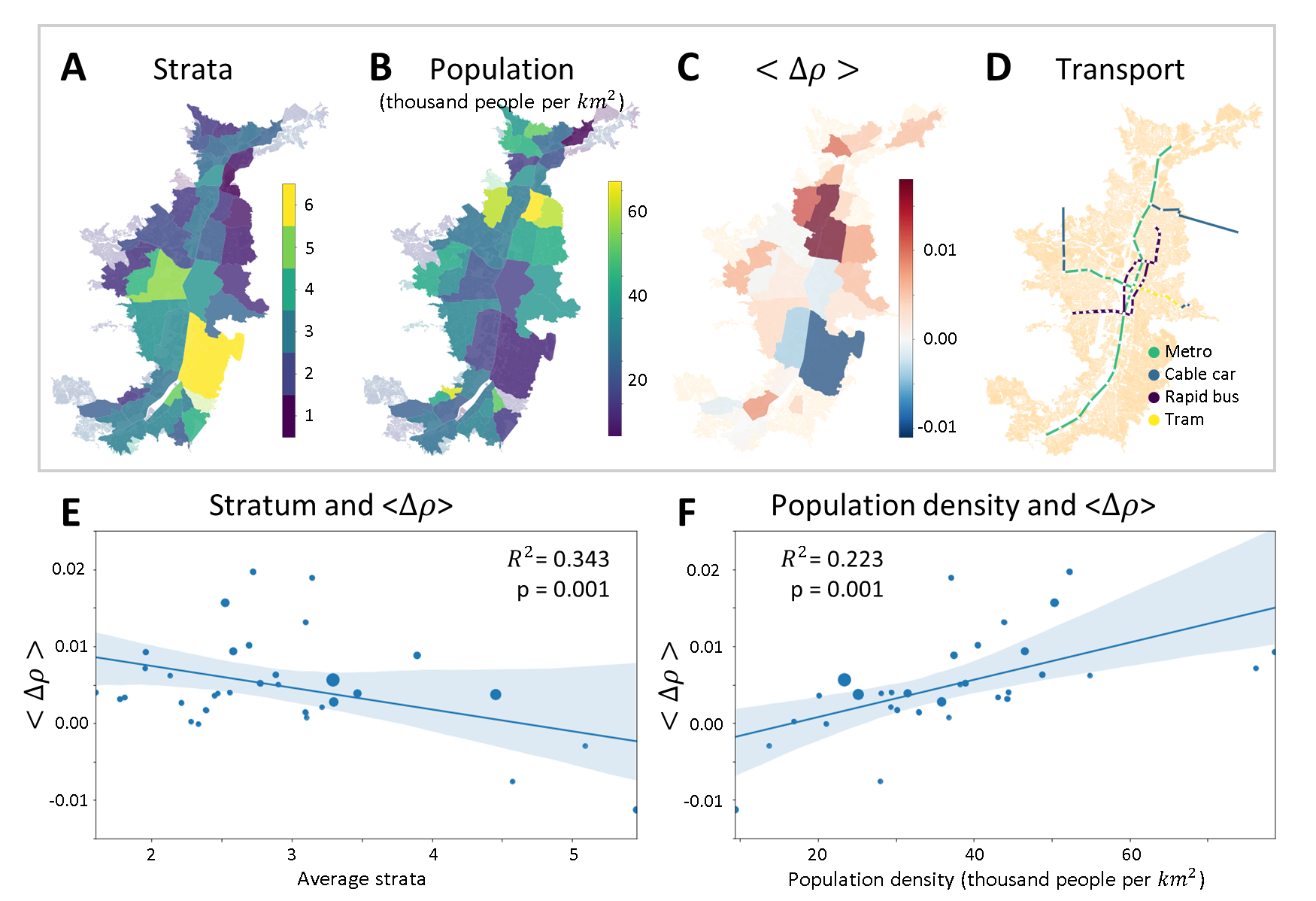}
  \caption{\textbf{Socio-economic strata and population density.} \textbf{(A)} The comunas of Medell\'in  according to their strata. Rich neighbourhoods (in yellow) are concentrated in El Poblado, while the outskirts usually belong to the poorest strata. \textbf{(B)} The most densely populated comunas are located just north of the city centre. \textbf{(C)} We show the average difference in density between visible and registered commercial firms, $<\Delta\rho>$, at the comuna level. Those in red have a relatively high density of visible firms, and those in blue have a larger concentration of commercial registered firms. \textbf{(D)} There is a significant negative relationship (p<0.001) between the average stratum of a comuna and $<\Delta\rho>$ at the comuna level. \textbf{(D)} Similarly, we find a significant relationship between the population density and $<\Delta\rho>$ at the comuna level.}
  \label{strata}
\end{figure*} 

\subsection*{Informal firms concentrate in poor but densely populated neighbourhoods with few complex industries}

The analysis above suggests that visible firms tend to concentrate in poorer areas away from the traditional economic center of the city. Here we are specifically interested in uncovering the density, socio-demographic status and industrial profile of neighbourhoods which are home to many visible firms but few registered firms. 

Granular data on socio-economic demographics is rarely available for developing cities. Here we exploit a policy of the Colombian government, which aims to progressively adjust charges for public utilities and services, in order to infer socio-economic status at a neighbourhood level. As a result of this policy, all neighbourhoods have been classified into six strata with 1 being the poorest and 6 the richest. Medell\'in has 10 municipalities that are subdivided into 66 comunas. Figure \ref{strata}A shows the mean stratum of each comuna. We observe that the richest stratum concentrates in El Poblado, south of the city centre, while the majority of the city belongs to strata 2-4. While strata do not perfectly correlate with income or other socio-economic variables, it has been widely used as a proxy for socio-economic status in the academic literature \cite{lotero2016rich, heroy2020controlling}. 

We show population density and $<\Delta \rho>$ at comuna level in Figure \ref{strata}B-C and the layout of the metro and bus rapid transport in Figure \ref{strata}D. We immediately observe that larger values of $\Delta \rho$ are associated with lower stratum (poorer areas) but larger population densities. Figure \ref{strata}E and F confirm that there is a statistically significant correlation between strata and population density with $\Delta\rho$ at the comuna level. Hence, poorer comunas with a high population density are home to a higher density of visible firms than registered commercial firms. Figure \ref{strata}D shows that some of these areas are also well served by the f\^eted metro system. 

Hence, it appears that informal commercial firms concentrate close to customers in poor but well-connected areas that are not being served by registered firms. This result is consistent with the idea that amenities and commercial firms will tend to locate close to consumers \cite{berry1988market, christaller1966central}, and that informal firms have lower barriers to entry that allow them source workers and meet demand in poorer areas \cite{la2014informality, o2018skill}. It is also backed up by research on South Africa which cited proximity and convenience as a key driver for customers of informal firms \cite{peberdy2018locating}. 

\begin{figure*}[t!]
  \includegraphics[width=\linewidth]{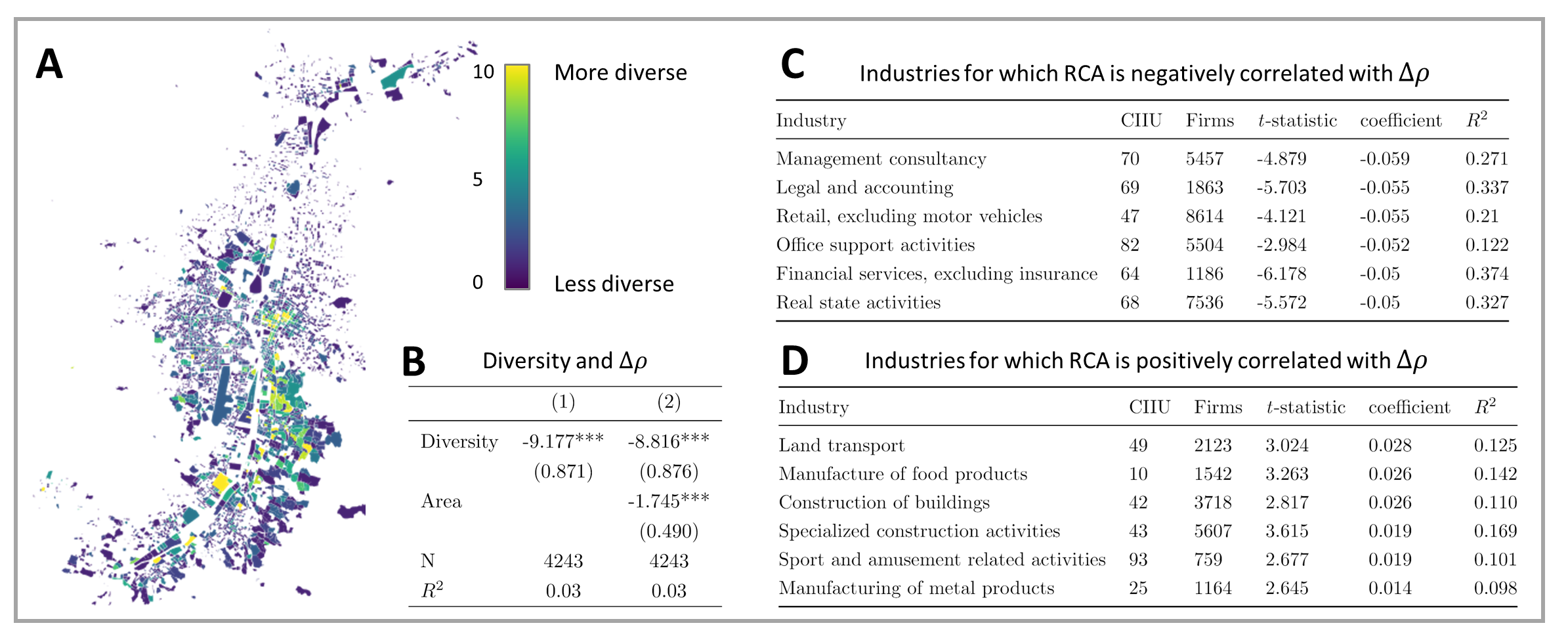}
  \caption{\textbf{Industrial structure.} \textbf{(A)} Map showing industrial diversity of neighbourhoods. \textbf{(B)} The relationship between industrial diversity and $\Delta\rho$ at the neighbourhood level. In the second column we control for the size of the neighbourhood. Less industrially diverse neighbourhoods are home to higher concentrations of visible firms relative to registered commercial firms. \textbf{(C-D)} For each industry sector (at the two digit level), we regress $\Delta\rho$ with RCA (the concentration of the sector) across comunas. We find concentration in complex sectors such as business and legal activities is negatively associated with $\Delta\rho$ (higher density of registered commercial firms) while concentration in low complexity activities such as manufacturing and construction is positively correlated with $\Delta\rho$.}
  \label{complexity}
\end{figure*}

Next we investigate the presence of visible and registered firms with respect to the local industrial profile. In particular, we would expect to find a higher density of informal firms in neighbourhoods with few sophisticated industries. Conversely, we would expect a higher density of registered firms in neighbourhoods that are home to complex industries such as finance, engineering or law. This is consistent with the argument that informal firms can been seen as those that require small teams with few specialised skills, while formal firms require larger teams of specialised skills found in large dense agglomerations \cite{o2018skill}. 

Building on an established literature that has shown that complex activities are located in industrially diverse places home to many skills and capabilities \cite{hidalgo2009building}, we construct a simple proxy for the industrial complexity of each neighbourhood by computing an industry diversity score. It is calculated for each neighbourhood by counting the number of distinct industries (at the 4-digit level) that are represented by at least one registered firm. These neighbourhoods are small geographical units consisting of a few blocks. Figure \ref{complexity}B shows that there is a statistically significant negative relationship between this score and $<\Delta\rho>$ at a neighbourhood level, confirming that less diverse neighbourhoods tend to be home to a higher density of visible firms relative to registered commercial firms. We show a comparable result at the comuna level in Appendix \ref{diver}.

Since diversity is an imperfect metric for the industrial sophistication of a neighbourhood, we investigate the distribution of visible firms with respect to individual sectors. To do this, for each comuna, we calculate the RCA (revealed comparative advantage from \cite{balassa1965trade}, see Materials and Methods), a metric capturing industry concentration in a place for each of the 88 industry sectors at the 2-digit level. In order to probe the relationship between the RCA and $<\Delta\rho>$ for each sector, we applied weighted regression at the comuna level (weighted by the number of firms in the comuna, we show the unweighted version in Appendix \ref{unweighted}). We show the top and bottom 6 sectors ranked by the size of the regression coefficient in Figure \ref{complexity}C-D. We find that comunas with a comparative advantage in complex sectors such as business and legal activities are negatively associated with $<\Delta\rho>$ (i.e., they have higher density of registered commercial firms than visible firms) while comunas concentrated in low complexity activities such as land transport, manufacturing and construction are positively correlated with $<\Delta\rho>$. Hence, visible firms tend to outnumber registered firms in these areas. 

\subsection*{Land use zoning is ineffective across all strata}

\begin{figure*}[t!]
  \includegraphics[width=\linewidth]{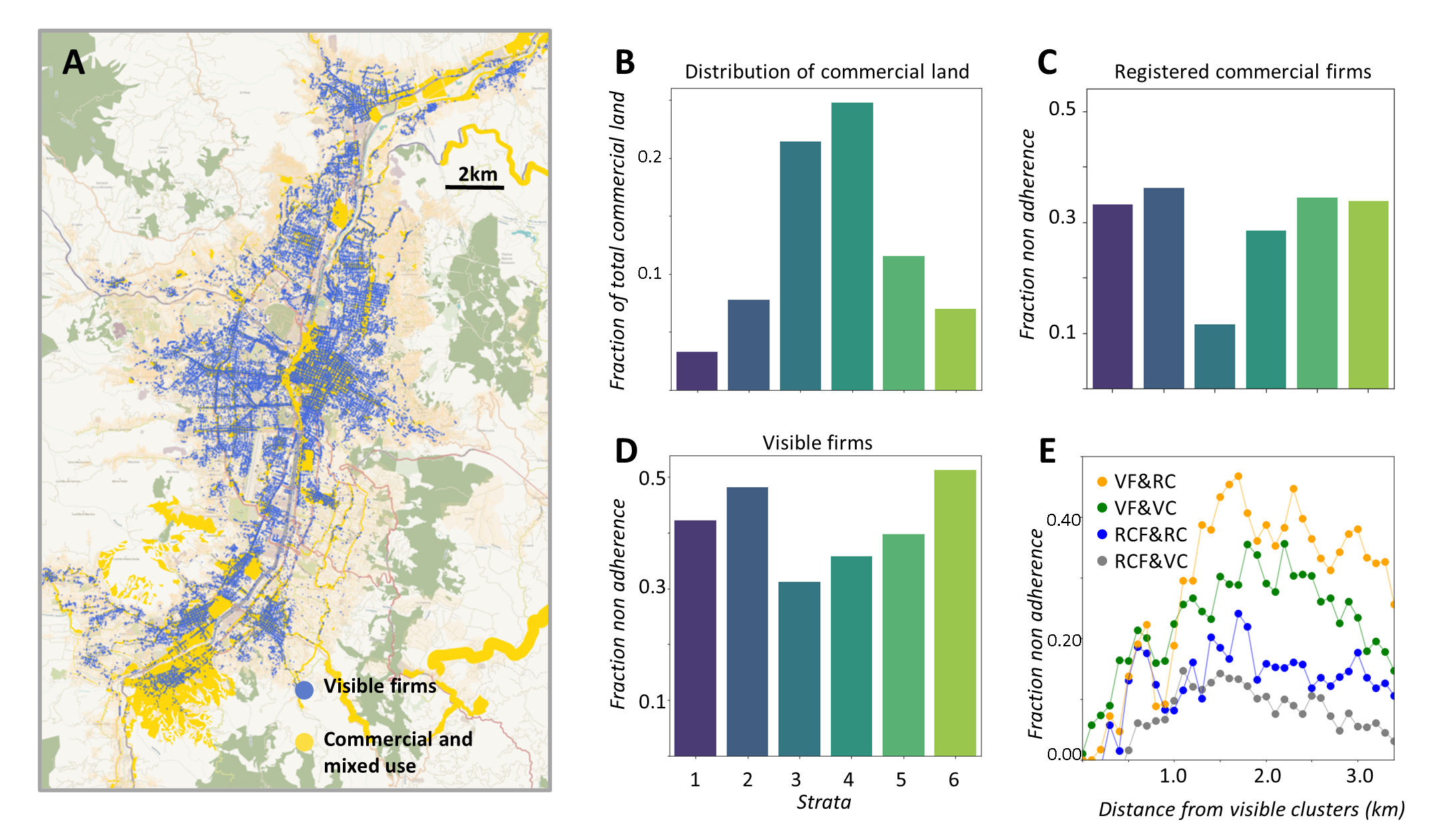}
  \caption{\textbf{Land use.} \textbf{(A)} Map of commercial land and visible firms. \textbf{(B)} Distribution of commercial or mixed use land across strata. \textbf{(C)} Non adherence across socio-economic strata for registered commercial firms. \textbf{(D)} Non adherence across socio-economic strata for visible firms. \textbf{(E)} Non-adherence of visible firms (VF) and registered commercial firms (RCF) as a function of distance from both visible clusters (VC) and formal clusters (FC). We find the highest level of non-adherence of visible firms occurs at around 2 km from the centre of the formal clusters, with a lower peak for non-adherence around visible clusters.}
  \label{land}
\end{figure*}

Zoning plans have been widely used as a tool for managing urban growth. But it has been pointed out that the enforcement is many times selective \cite{bartram2019going}, that it can reinforce inequality and related dynamics \cite{lyles2016local, bartram2019going}, and that there is generally a large amount of non conformance to the plans \cite{shen2020does}. Problematically, non adherence to zoning restrictions is usually associated to poorer neighbourhoods further complicating the debate \cite{ghertner2012nuisance}. Furthermore, work on street commerce and land use suggests that such firms benefit from mixed use zoning, enabling firms to flexibly locate near consumers rather than designated shopping zones \cite{sevtsuk2020street}. Here we investigate adherence to zoning of visible and registered firms by comparing their location to the official land use plan of the city (Appendix \ref{data}) as seen in Figure \ref{land}A. 

We investigate the adherence of firms through the lens of socio-economic segregation or stratum. In Figure \ref{land}B we show that most of the commercial and mixed-use land belongs to the middle strata 3 and 4. We define the non-adherence rate as the fraction of firms in non mixed or commercial zones as a share of total firms. Consistent with Figure \ref{land}B, we find that non-adherence of registered commercial firms is lower in the middle strata but higher in both poor and wealthy areas (with less available commercial and mixed use land). Similar patterns are found for the larger set of registered firms, see Appendix \ref{landuseReg}. Contrary to common perceptions, however, Figure \ref{land}C shows that non-adherence of visible firms is reasonably steady across all strata. We find that for all six strata between 30\% and 50\% of  visible firms are located on non commercial land. Furthermore, the richest stratum exhibits the highest level of non-adherence, while strata 3 and 4 have the lowest level. Hence, even in wealthy areas the incentives for commercial firms to locate are stronger that the enforcement of zoning regulations. Overall, adherence to zoning regulations is low for both visible and registered firms across all strata including wealthy areas.

We found above that visible firms tend to concentrate (relative to registered commercial firms) on the outskirts of formal clusters. Here we investigate the levels of non-adherence of firms relative to their distance to a cluster. Figure \ref{land}D plots the average non-adherence of visible firms and registered commercial firms against the distance to the midpoint of the closest formal cluster or visible cluster (here the latter set excludes the overlapping cluster, Centro). We find the highest level of non-adherence of visible firms occurs at around 2 km from the center of formal clusters, with a lower peak for non-adherence around visible clusters. Hence it appears that visible firms, which include informal commercial firms, are most likely to violate zoning laws when located close to formal centers, likely benefiting from both a combination of a dense customer base and linkages with formal firms. We also consider the non-adherence of registered commercial firms, finding a smaller peak at a similar distance from formal clusters. Finally, consistent with what we would expect, the lowest level of non-adherence with a minimal peak is found for registered commercial firms around visible clusters. 

\section*{Discussion}

This paper proposes a new methodology for identifying and tracking commercial activity in informal cities using street view imagery. This approach is a fast and cost effective alternative to surveys and commercial registries. By focusing on the metropolitan of Medell\'in we show that the detection algorithm allows us to map the spatial distribution of visible commercial activity and identify economic clusters with a high density of visible firms. Comparing our dataset to the set of registered firms, we demonstrate that we capture activity that is not reflected in the official records, particularly in poorer and more densely populated regions of the metropolitan area.   

Our methodology is not without limitations. Firstly, the set of visible firms is a specific subset of all firms, as it only includes those that are easily identifiable at street level. Visible firms include retail activities, personal services and other similar activities and amenities. While these are arguably some of the most dynamic and informal sectors of the economy, and hence the dataset is thus particularly useful for capturing economic activity in a developing city, any analysis done on this set of firms must take this into account. Secondly, while comparable to related efforts to detect shopfronts \cite{yu2015large, ye2019urban, sharifi2020detecting}, the algorithm does not perfectly identify commercial firms. Specifically, while these algorithms show great precision in identifying firms, their recall is not as high, which means they are likely to underestimate the number of firms. Thirdly, although the methodology is easily transferable to other contexts where street imagery is available, it does require training data for the detector. In our case, this involved many hours of manually labelling imagery. Future work will investigate the extent to which new regions require a bespoke training set, or whether images trained on one city can be used to identify facades in another. 

There are many other avenues for future work. Here were have considered imagery for just one year, but analysis of imagery over longer time periods could provide important information about the evolution of the spatial concentration of economic activity over time, and the impact, for example, of public transport and road investments. We cannot easily disentangle formal from informal firms in our dataset, and use a registry of formal firms in order to identify areas with an 'excess' concentration of visible firms relative to registered firms in order to infer the presence of informal firms. Future work might aim to further match these datasets, or deploy other techniques - such as training the detector to identify informal firms - in order further distinguish informal from formal firms. Finally, there are other possible approaches to close the gaps in official data, such as crowd-sourced data. Google Street Maps and Open Street Maps, for example, provide information on amenities such as bars, restaurants and shops. While crowd-sourced data is susceptible to self selection and other biases \cite{quattrone2015there}, a potential avenue for future research would be to integrate and benchmark against these other sources.


\section*{Materials and Methods}


\subsection*{Density estimation}

KDE was applied to the detections in the region of Medell\'in. A grid of 200 m by 200 m cells was drawn over the extension of the city. The level of smoothing is dictated by the bandwidth which we fixed at 150m, which is equivalent to walking two blocs (see \cite{fotheringham2000quantitative}). For each cell in the grid we obtain a density value, and we re-scale these values so that they sum up to one.

\subsection*{Clustering} 

To identify the clusters of visible firms, we followed the methodology of \cite{arribas2014validity} and applied LISA \cite{anselin1995local} to our dataset of visible firms. This method calculates a local version of the traditional Moran I auto-correlation statistic. Contiguous cells that show auto-correlation above a certain significance level are grouped into clusters, a p-value of 0.10 is the usual choice in the literature \cite{arribas2014validity}. To investigate the persistence of the clusters, we vary the minimum density required for a cell to be included. The clusters found at each different threshold form a family of nested clusters described by the dendrogram in Figure \ref{clusters}.

\subsection*{Socio-economic strata in Medell\'in} 

The national census of Colombia \cite{DANE} provides the official socio-economic stratum at the neighbourhood level. We focus our analysis on the urban region of Medell\'in, as defined in the census. This region spans 10 municipalities. Municipalities are divided into 66 comunas (or \textit{macrozonas}) which can be further divided into neighbourhoods. The majority of our analysis has been carried at the comuna level, with the instances in which we use neighbourhoods clearly indicated.

\subsection*{Revealed comparative advantage}

The RCA \cite{balassa1965trade}, which we calculated at the comuna level, is given by 
\[
    RCA_{i,c} = \frac{|F_c \cap F_i|/| F_i|}{|F_i|/|F|}
\]
where $F_c$ is the set of registered firms in comuna $c$, $F_i$ is the set of registered firms in industry $i$, and $F$ is the set of all registered firms. If it is bigger than one it shows that that industry is more concentrated in that comuna than it is in the whole region.

\subsection*{Land use and visible firms} 

Land use in Medell\'in is governed by the POT \cite{POT}. This plan allocates a fraction of the total land to commercial and mixed-use; theoretically, all firms should be located inside these areas. Using the geo-location of the detected firms we verified which firms were located in this region. We labelled every detection according to the socio-economic stratum in which it falls. For each stratum $i$ we calculated the following measure of (non-) adherence to land use:
\[
A(s) = 1 -\frac{|C_s \cap C_A|}{|C_A|}  
\]
where $C_s$ is the set of visible firms that belong to stratum $s$ and $C_A$ is the set of visible firms that are located in commercial and mixed used land. 

We repeated this analysis for the set of registered firms, and for commercial registered firms, which we obtained from the industries classified as street commerce in \cite{sevtsuk2020street}. .


\bibliography{visibleLit}

\begin{thebibliography}{10}

\bibitem{un2018}
UN, World urbanization prospects, {\it Tech. rep.\/}, UN (2018).

\bibitem{rosenthal2004evidence}
S.~S. Rosenthal, W.~C. Strange, Evidence on the nature and sources of
  agglomeration economies.
\newblock {\it Handbook of regional and urban economics\/} (Elsevier, 2004),
  vol.~4, pp. 2119--2171.

\bibitem{duranton2005testing}
G.~Duranton, H.~G. Overman, Testing for localization using micro-geographic
  data.
\newblock {\it The Review of Economic Studies\/} {\bf 72}, 1077--1106 (2005).

\bibitem{bryan2020cities}
G.~Bryan, E.~Glaeser, N.~Tsivanidis, Cities in the developing world  (2020).

\bibitem{goswami2019jobs}
A.~G. Goswami, S.~V. Lall, Jobs and land use within cities: A survey of theory,
  evidence, and policy.
\newblock {\it The World Bank Research Observer\/} {\bf 34}, 198--238 (2019).

\bibitem{ilo2018}
International-Labour-Organization, Women and men in the informal economy : a
  statistical picture, {\it Tech. rep.\/}, International-Labour-Organization,
  Geneva, Switzerland (2018).

\bibitem{worldbank}
{World Bank}, Enterprise surveys,
  http://www.enterprisesurveys.org/Data/ExploreTopics/informality (2017).

\bibitem{henley2009defining}
A.~Henley, G.~R. Arabsheibani, F.~G. Carneiro, On defining and measuring the
  informal sector: Evidence from brazil.
\newblock {\it World Development\/} {\bf 37}, 992--1003 (2009).

\bibitem{henderson2012measuring}
J.~V. Henderson, A.~Storeygard, D.~N. Weil, Measuring economic growth from
  outer space.
\newblock {\it American Economic Review\/} {\bf 102}, 994--1028 (2012).

\bibitem{crooks2015crowdsourcing}
A.~Crooks, D.~Pfoser, A.~Jenkins, A.~Croitoru, A.~Stefanidis, D.~Smith,
  S.~Karagiorgou, A.~Efentakis, G.~Lamprianidis, Crowdsourcing urban form and
  function.
\newblock {\it International Journal of Geographical Information Science\/}
  {\bf 29}, 720--741 (2015).

\bibitem{boeing2021spatial}
G.~Boeing, Spatial information and the legibility of urban form: Big data in
  urban morphology.
\newblock {\it International Journal of Information Management\/} {\bf 56},
  102013 (2021).

\bibitem{henderson2020building}
J.~V. Henderson, T.~Regan, A.~J. Venables, Building the city: from slums to a
  modern metropolis.
\newblock {\it The Review of Economic Studies\/}  (2020).

\bibitem{kreindler2021measuring}
G.~E. Kreindler, Y.~Miyauchi, Measuring commuting and economic activity inside
  cities with cell phone records, {\it Tech. rep.\/}, National Bureau of
  Economic Research (2021).

\bibitem{jean2016combining}
N.~Jean, M.~Burke, M.~Xie, W.~M. Davis, D.~B. Lobell, S.~Ermon, Combining
  satellite imagery and machine learning to predict poverty.
\newblock {\it Science\/} {\bf 353}, 790--794 (2016).

\bibitem{goldblatt2019can}
R.~Goldblatt, K.~Heilmann, Y.~Vaizman, {\it Can medium-resolution satellite
  imagery measure economic activity at small geographies? Evidence from Landsat
  in Vietnam\/} (The World Bank, 2019).

\bibitem{kuffer2016slums}
M.~Kuffer, K.~Pfeffer, R.~Sliuzas, Slums from space—15 years of slum mapping
  using remote sensing.
\newblock {\it Remote Sensing\/} {\bf 8}, 455 (2016).

\bibitem{duque2015measuring}
J.~C. Duque, J.~E. Patino, L.~A. Ruiz, J.~E. Pardo-Pascual, Measuring
  intra-urban poverty using land cover and texture metrics derived from remote
  sensing data.
\newblock {\it Landscape and Urban Planning\/} {\bf 135}, 11--21 (2015).

\bibitem{gebru2017using}
T.~Gebru, J.~Krause, Y.~Wang, D.~Chen, J.~Deng, E.~L. Aiden, L.~Fei-Fei, Using
  deep learning and google street view to estimate the demographic makeup of
  neighborhoods across the united states.
\newblock {\it Proceedings of the National Academy of Sciences\/} {\bf 114},
  13108--13113 (2017).

\bibitem{naik2017computer}
N.~Naik, S.~D. Kominers, R.~Raskar, E.~L. Glaeser, C.~A. Hidalgo, Computer
  vision uncovers predictors of physical urban change.
\newblock {\it Proceedings of the National Academy of Sciences\/} {\bf 114},
  7571--7576 (2017).

\bibitem{gonzalez2020automatic}
D.~Gonzalez, D.~Rueda-Plata, A.~B. Acevedo, J.~C. Duque, R.~Ramos-Pollan,
  A.~Betancourt, S.~Garcia, Automatic detection of building typology using deep
  learning methods on street level images.
\newblock {\it Building and Environment\/} {\bf 177}, 106805 (2020).

\bibitem{rueda2021use}
D.~Rueda-Plata, D.~Gonz{\'a}lez, A.~Acevedo, J.~Duque, R.~Ramos-Poll{\'a}n, Use
  of deep learning models in street-level images to classify one-story
  unreinforced masonry buildings based on roof diaphragms.
\newblock {\it Building and Environment\/} {\bf 189}, 107517 (2021).

\bibitem{sevtsuk2020street}
A.~Sevtsuk, {\it Street commerce: Creating vibrant urban sidewalks\/}
  (University of Pennsylvania Press, 2020).

\bibitem{berry1993challenges}
B.~J. Berry, H.-M. Kim, Challenges to the monocentric model.
\newblock {\it Geographical Analysis\/} {\bf 25}, 1--4 (1993).

\bibitem{gordon1986distribution}
P.~Gordon, H.~W. Richardson, H.~L. Wong, The distribution of population and
  employment in a polycentric city: the case of {L}os {A}ngeles.
\newblock {\it Environment and Planning A\/} {\bf 18}, 161--173 (1986).

\bibitem{giuliano1991subcenters}
G.~Giuliano, K.~A. Small, Subcenters in the los angeles region.
\newblock {\it Regional Science and Urban economics\/} {\bf 21}, 163--182
  (1991).

\bibitem{gordon1996beyond}
P.~Gordon, H.~W. Richardson, Beyond polycentricity: the dispersed metropolis,
  los angeles, 1970-1990.
\newblock {\it Journal of the American Planning Association\/} {\bf 62},
  289--295 (1996).

\bibitem{lucas2002internal}
R.~E. Lucas, E.~Rossi-Hansberg, On the internal structure of cities.
\newblock {\it Econometrica\/} {\bf 70}, 1445--1476 (2002).

\bibitem{ahlfeldt2015economics}
G.~M. Ahlfeldt, S.~J. Redding, D.~M. Sturm, N.~Wolf, The economics of density:
  Evidence from the {B}erlin wall.
\newblock {\it Econometrica\/} {\bf 83}, 2127--2189 (2015).

\bibitem{mills1980comparison}
E.~S. Mills, J.~P. Tan, A comparison of urban population density functions in
  developed and developing countries.
\newblock {\it Urban Studies\/} {\bf 17}, 313--321 (1980).

\bibitem{bertaud2003spatial}
A.~Bertaud, S.~Malpezzi, The spatial distribution of population in 48 world
  cities: Implications for economies in transition.
\newblock {\it Center for urban land economics research, University of
  Wisconsin\/} {\bf 32}, 54--55 (2003).

\bibitem{arribas2014validity}
D.~Arribas-Bel, F.~Sanz-Gracia, The validity of the monocentric city model in a
  polycentric age: Us metropolitan areas in 1990, 2000 and 2010.
\newblock {\it Urban Geography\/} {\bf 35}, 980--997 (2014).

\bibitem{marshall1920economics}
A.~Marshall, {\it The economics of industry\/} (Macmillan and Company, 1920).

\bibitem{krugman1991increasing}
P.~Krugman, Increasing returns and economic geography.
\newblock {\it Journal of Political Economy\/} {\bf 99}, 483--499 (1991).

\bibitem{melo2009meta}
P.~C. Melo, D.~J. Graham, R.~B. Noland, A meta-analysis of estimates of urban
  agglomeration economies.
\newblock {\it Regional Science and Urban Economics\/} {\bf 39}, 332--342
  (2009).

\bibitem{kolko2010urbanization}
J.~Kolko, Urbanization, agglomeration, and coagglomeration of service
  industries.
\newblock {\it Agglomeration economics\/} (University of Chicago Press, 2010),
  pp. 151--180.

\bibitem{diodato2018industries}
D.~Diodato, F.~Neffke, N.~O’Clery, Why do industries coagglomerate? how
  marshallian externalities differ by industry and have evolved over time.
\newblock {\it Journal of Urban Economics\/} {\bf 106}, 1--26 (2018).

\bibitem{andersson2019economic}
M.~Andersson, J.~P. Larsson, J.~Wernberg, The economic microgeography of
  diversity and specialization externalities--firm-level evidence from swedish
  cities.
\newblock {\it Research Policy\/} {\bf 48}, 1385--1398 (2019).

\bibitem{rosenthal2003geography}
S.~S. Rosenthal, W.~C. Strange, Geography, industrial organization, and
  agglomeration.
\newblock {\it Review of Economics and Statistics\/} {\bf 85}, 377--393 (2003).

\bibitem{bartik1987urban}
T.~J. Bartik, V.~K. Smith, Urban amenities and public policy.
\newblock {\it Handbook of Regional and Urban Economics\/} (Elsevier, 1987),
  vol.~2, pp. 1207--1254.

\bibitem{albouy2016cities}
D.~Albouy, What are cities worth? {L}and rents, local productivity, and the
  total value of amenities.
\newblock {\it Review of Economics and Statistics\/} {\bf 98}, 477--487 (2016).

\bibitem{berry1988market}
B.~J.~L. Berry, J.~B. Parr, {\it Market centers and retail location\/}
  (Prentice Hall, 1988).

\bibitem{glaeser2001consumer}
E.~L. Glaeser, J.~Kolko, A.~Saiz, Consumer city.
\newblock {\it Journal of Economic Geography\/} {\bf 1}, 27--50 (2001).

\bibitem{christaller1966central}
W.~Christaller, {\it Central places in southern Germany\/}, vol.~10
  (Prentice-Hall, 1966).

\bibitem{meijers2007central}
E.~Meijers, From central place to network model: theory and evidence of a
  paradigm change.
\newblock {\it Tijdschrift voor economische en sociale geografie\/} {\bf 98},
  245--259 (2007).

\bibitem{wang2020polycentric}
M.~Wang, Polycentric urban development and urban amenities: Evidence from
  chinese cities.
\newblock {\it Environment and Planning B: Urban Analytics and City Science\/}
  p. 2399808320951205 (2020).

\bibitem{overman2005cities}
H.~G. Overman, A.~J. Venables, {\it Cities in the developing world\/}, no. 695
  (Centre for Economic Performance, London School of Economics and
  Political~…, 2005).

\bibitem{moreno2012critical}
A.~Moreno-Monroy, Critical commentary. informality in space: Understanding
  agglomeration economies during economic development.
\newblock {\it Urban Studies\/} {\bf 49}, 2019--2030 (2012).

\bibitem{glaeser2017urban}
E.~Glaeser, J.~V. Henderson, Urban economics for the developing world: An
  introduction.
\newblock {\it Journal of Urban Economics\/} {\bf 98}, 1--5 (2017).

\bibitem{krygsman2016cape}
S.~Krygsman, T.~de~Jong, O.~Verkoren, Cape town and its employment centres:
  Monocentric, polycentric or somewhere in-between?
\newblock {\it South African Geographers\/} p.~66 (2016).

\bibitem{garcia2019agglomeration}
G.~A. Garc{\'\i}a, Agglomeration economies in the presence of an informal
  sector: the colombian case.
\newblock {\it Revue dEconomie Regionale Urbaine\/} pp. 355--388 (2019).

\bibitem{tanaka2020agglomeration}
K.~Tanaka, Y.~Hashiguchi, Agglomeration economies in the formal and informal
  sectors: a bayesian spatial approach.
\newblock {\it Journal of Economic Geography\/} {\bf 20}, 37--66 (2020).

\bibitem{duranton2009cities}
G.~Duranton, Are cities engines of growth and prosperity for developing
  countries?
\newblock {\it Urbanization and Growth\/} pp. 67--114 (2009).

\bibitem{giuliani2005cluster}
E.~Giuliani, Cluster absorptive capacity: why do some clusters forge ahead and
  others lag behind?
\newblock {\it European Urban and Regional Studies\/} {\bf 12}, 269--288
  (2005).

\bibitem{gebreeyesus2013innovation}
M.~Gebreeyesus, P.~Mohnen, Innovation performance and embeddedness in networks:
  evidence from the {E}thiopian footwear cluster.
\newblock {\it World Development\/} {\bf 41}, 302--316 (2013).

\bibitem{mukim2015coagglomeration}
M.~Mukim, Coagglomeration of formal and informal industry: evidence from india.
\newblock {\it Journal of Economic Geography\/} {\bf 15}, 329--351 (2015).

\bibitem{fernandez2018informalidad}
C.~Fernandez, {\it et~al.\/}, Informalidad empresarial en colombia, {\it Tech.
  rep.\/}, Fedesarrollo (2018).

\bibitem{o2018skill}
N.~O’Clery, J.~C. Chaparro, A.~Gomez-Lievano, E.~Lora, Skill diversity as the
  foundation of formal employment creation in cities, {\it Tech. rep.\/},
  Technical report, Working Paper at Center for International Development
  at~… (2018).

\bibitem{lotero2016rich}
L.~Lotero, R.~G. Hurtado, L.~M. Flor{\'\i}a, J.~G{\'o}mez-Garde{\~n}es, Rich do
  not rise early: spatio-temporal patterns in the mobility networks of
  different socio-economic classes.
\newblock {\it Royal Society Open Science\/} {\bf 3}, 150654 (2016).

\bibitem{arbaux2012Medellin}
M.~H. Arbaux, A.~E. Restrepo, J.~Giraldo, {\it Medell{\'\i}n: environment
  urbanism society\/} (Universidad EAFIT, 2012).

\bibitem{duque2013localizacion}
V.~G. Duque, Localizaci{\'o}n espacial de la actividad econ{\'o}mica en
  medell{\'i}n, 2005--2010 un enfoque de econom{\'\i}a urbana.
\newblock {\it Ensayos sobre pol{\'i}tica econ{\'o}mica\/} {\bf 31}, 215--266
  (2013).

\bibitem{shorten2019survey}
C.~Shorten, T.~M. Khoshgoftaar, A survey on image data augmentation for deep
  learning.
\newblock {\it Journal of Big Data\/} {\bf 6}, 60 (2019).

\bibitem{ren2015faster}
S.~Ren, K.~He, R.~Girshick, J.~Sun, {\it Advances in neural information
  processing systems\/} (2015), pp. 91--99.

\bibitem{yu2015large}
Q.~Yu, C.~Szegedy, M.~C. Stumpe, L.~Yatziv, V.~Shet, J.~Ibarz, S.~Arnoud, Large
  scale business discovery from street level imagery.
\newblock {\it arXiv preprint arXiv:1512.05430\/}  (2015).

\bibitem{ye2019urban}
N.~Ye, B.~Wang, M.~Kita, M.~Xie, W.~Cai, Urban commerce distribution analysis
  based on street view and deep learning.
\newblock {\it IEEE Access\/} {\bf 7}, 162841--162849 (2019).

\bibitem{sharifi2020detecting}
S.~Sharifi~Noorian, S.~Qiu, A.~Psyllidis, A.~Bozzon, G.-J. Houben, {\it
  Proceedings of the 2020 International Conference on Multimedia Retrieval\/}
  (2020), pp. 495--501.

\bibitem{rosenblatt1956remarks}
M.~Rosenblatt, Remarks on some nonparametric estimates of a density function.
\newblock {\it Annals of Mathematical Statistics\/}  (1956).

\bibitem{anselin1995local}
L.~Anselin, Local indicators of spatial association—lisa.
\newblock {\it Geographical Analysis\/} {\bf 27}, 93--115 (1995).

\bibitem{dominguez2019agglomeration}
A.~Dominguez, Agglomeration effects and informal firms in the internal
  structure of cities.
\newblock {\it Applied Economic Analysis\/}  (2019).

\bibitem{heroy2020controlling}
S.~Heroy, I.~Loaiza, A.~Pentland, N.~O'Clery, Controlling covid-19: Labor
  structure is more important than lockdown policy.
\newblock {\it Journal of the Royal Society Interface\/} {\bf 18}, 20201035
  (2021).

\bibitem{la2014informality}
R.~La~Porta, A.~Shleifer, Informality and development.
\newblock {\it Journal of Economic Perspectives\/} {\bf 28}, 109--26 (2014).

\bibitem{peberdy2018locating}
S.~Peberdy, Locating the informal sector in the gauteng city-region and beyond.
\newblock {\it The Changing Space Economy of City-Regions\/} (Springer, 2018),
  pp. 185--211.

\bibitem{hidalgo2009building}
C.~A. Hidalgo, R.~Hausmann, The building blocks of economic complexity.
\newblock {\it Proceedings of the national academy of sciences\/} {\bf 106},
  10570--10575 (2009).

\bibitem{balassa1965trade}
B.~Balassa, Trade liberalisation and ``revealed'' comparative advantage1.
\newblock {\it The Manchester School\/} {\bf 33}, 99--123 (1965).

\bibitem{bartram2019going}
R.~Bartram, Going easy and going after: Building inspections and the selective
  allocation of code violations.
\newblock {\it City \& Community\/} {\bf 18}, 594--617 (2019).

\bibitem{lyles2016local}
W.~Lyles, P.~Berke, G.~Smith, Local plan implementation: Assessing conformance
  and influence of local plans in the united states.
\newblock {\it Environment and Planning B: Planning and Design\/} {\bf 43},
  381--400 (2016).

\bibitem{shen2020does}
X.~Shen, X.~Wang, Z.~Zhang, L.~Fei, Does non-conforming urban development mean
  the failure of zoning? a framework for conformance-based evaluation.
\newblock {\it Environment and Planning B: Urban Analytics and City Science\/}
  p. 2399808320926179 (2020).

\bibitem{ghertner2012nuisance}
D.~A. Ghertner, Nuisance talk and the propriety of property: middle class
  discourses of a slum-free delhi.
\newblock {\it Antipode\/} {\bf 44}, 1161--1187 (2012).

\bibitem{quattrone2015there}
G.~Quattrone, L.~Capra, P.~De~Meo, {\it Proceedings of the 18th ACM Conference
  on Computer Supported Cooperative Work \& Social Computing\/} (2015), pp.
  1021--1032.

\bibitem{fotheringham2000quantitative}
A.~S. Fotheringham, C.~Brunsdon, M.~Charlton, {\it Quantitative geography:
  perspectives on spatial data analysis\/} (Sage, 2000).

\bibitem{DANE}
{Departamento Administrativo Nacional de Estadística (DANE)}, Censo nacional
  de población y vivienda,
  https://www.dane.gov.co/index.php/en/estadisticas-por-tema/demografia-y-poblacion/censo-nacional-de-poblacion-y-vivenda-2018
  (2020).

\bibitem{POT}
{Alcald\'ia de Medellin}, Plan de ordenamiento territorial,
  https://geomedellin-m-medellin.opendata.arcgis.com/datasets/gdb-pot-acuerdo48-de-2014
  (2014).

\bibitem{liu2016ssd}
W.~Liu, D.~Anguelov, D.~Erhan, C.~Szegedy, S.~Reed, C.-Y. Fu, A.~C. Berg, {\it
  European conference on computer vision\/} (Springer, 2016), pp. 21--37.

\bibitem{redmon2016you}
J.~Redmon, S.~Divvala, R.~Girshick, A.~Farhadi, {\it Proceedings of the IEEE
  conference on computer vision and pattern recognition\/} (2016), pp.
  779--788.

\bibitem{pan2009survey}
S.~J. Pan, Q.~Yang, A survey on transfer learning.
\newblock {\it IEEE Transactions on knowledge and data engineering\/} {\bf 22},
  1345--1359 (2009).

\end{thebibliography}

\bibliographystyle{ScienceAdvances}

\section*{Acknowledgments}

We would like to thank Samira Barzin, Samuel Heroy and Niclas Moneke for their feedback on the manuscript.

This article was completed with support from the the PEAK Urban Program, supported by UKRI’s Global Challenge Research Fund, Grant Ref: ES/P011055/1. 



\appendix
\vspace{1cm}
{\Large \bf \noindent Appendix} 

\section{Data}\label{data}

For our analysis, we have used three publicly available datasets, two of which are hosted by the Colombian statistics agency DANE (Departamento Administrativo Nacional de Estadística). 

The two DANE datasets are:
\begin{itemize}
\item The 2018 census (Censo nacional de población y vivienda)
www.dane.gov.co/ index.php/en/estadisticas-por-tema/demografia-y-poblacion/censo-nacional- de-poblacion-y-vivenda-2018
\item The 2018 firm registry (Directorio Estad\'istico de Empresas) https://www.dane.gov.co/ index.php/servicios-al-ciudadano/servicios-informacion/ directorio-estadistico/directorio-estadistico-de-empresas
\end{itemize}

The 2014 land use plan (Plan de ordenamiento territorial -POT) is hosted by the Medell\'in municipality: www.geomedellin-m-medellin.opendata.arcgis.com/datasets/ gdb-pot-acuerdo48-de-2014



To avoid potential identification of individual firms, the firms locations have been grouped into cells of 200m by 200m. The dataset consists of a grid covering the city of Medell\'in with the number of firms detected in each cell.

\section{Training the detector}\label{detector}

To create a training set for our detection algorithm, we randomly sampled $2,000$ points from the street network of Medell\'in, which we sourced from Open Street Maps (OSM) (Figure \ref{commerce}). By calling the Google street view API we obtained a panoramic image for each of these points, which we then transformed into two standard images, one for each side of the road. These images were manually labelled by drawing a bounding box around every facade and then labelling each facade as either commercial or non-commercial. The labelled dataset contains $\sim 2,000$ commercial facades and $\sim 6,000$ non-commercial. The images were randomly split into training (60\%), validation (20\%) and test (20\%) sets. The training set was then enriched following standard data augmentation procedures such as rotation, translation and noise addition \cite{shorten2019survey} which are common in detection and classification tasks.

Three architectures were considered for the detector: Single Shot Detection (SSD) \cite{liu2016ssd}, you only look once (YOLO) \cite{redmon2016you} and Faster R-CNN \cite{ren2015faster}. These detectors take an image as the input and output the location(s) of the object(s) of interest within the image, which in our case are the facades of commercial firms. Given that we only had a small training set, we used transfer learning \cite{pan2009survey}, which allowed us to retrain models that had been previously fitted for a similar object detection task. The three different architectures were trained using our training and validation sets, and their performance was compared using the test set. 

In our application we are mainly using the number of visible firms detected per image. Therefore, on top of evaluating the precision and recall of the detector, we defined two metrics to better reflect this objective. The first one, $C_f$, calculates the fraction of visible firms that we detect only on the images that contain visible firms. The second one, $Err_0$, calculates how many detections we have in images that contain no visible firms. 

$$
C_f:=\frac{1}{n_{r}}\sum_{i}^{n_{r}}\frac{\hat{c}_i}{c_{i}}  ,  \quad c_{i} >0  
$$

$$
Err_0:=\frac{1}{n_{nr}}\sum_{i}^{n_{nr}}{\hat{c}_i},  \quad  c_{i} =0
$$

Where:
\begin{itemize}
    \item $c_{i}$: number of visible firms in the $i$-th image. 
    \item $\hat{c}_i$: number of detections in the $i$-th image. 
    \item $n_{r}$: number of images in the dataset with at least one visible firm. 
    \item $n_{nr}$: number of images in the dataset with no visible firms. 
\end{itemize}

A well calibrated algorithm will score close to $1$ in the first one and close to $0$ in the second one.

\begin{center}
    \begin{tabular}{||c c c c||} 
        \hline
        Score & Training & Validation & Test \\ [0.5ex] 
        \hline
        $C_f$ & 0.985 & 1.011 & 0.955 \\ 
        \hline
        $Err_0$ & 0.034 & 0.222 & 0.200 \\[1ex] 
        \hline
    \end{tabular}
\\
    \begin{small} \textbf{Table 2.} Alternative metrics. We have calculated these metrics to evaluate the detector with respect to the number of visible firms per image.  \end{small}
   
\end{center}

As well as training the detector, it is necessary to identify the locations to which it will be applied in order fully cover the urban area of Medell\'in. To do so, we needed to carefully identify the points from which we should source the panoramic images. The street network of the Medell\'in metropolitan region was obtained from OSM, an open source project that freely provides road networks and other useful geographical information.

We chose points in the street network in such a way that the images would not overlap while also capturing (almost) all the facades. Since after processing a 360 panoramic image we obtain roughly 20 meters of each side of the street, we wanted to ensure that along each road segment we had an image every 20 meters. To do so, we first took all of the network crossings. For each segment of the network between two crossings we added the maximum number of points we could while ensuring they remained at least 20 meters apart. This heuristic allowed us to quickly create a set of points that gives us close to maximal coverage of the city. We then used Google's street view API to obtain a panoramic image for each point. Finally, we applied the detector to each image to produce a list of geo-referenced points with counts of detected visible firms.

Lastly, we investigated the impact of the detector's recall on the spatial distribution of the detections (visible firms). By sampling a random point and a radius we create a random region. We then calculate the ratio $R_r$ between the total number of firms and the number of detections in this region. If this ratio remain stable across different regions, then we know that the visible firms must have a very similar spatial distribution to the detections. 

$$
R_r:=\frac{\hat{c}_r}{c_{r}}
$$

We calculated the above ratio for 1000 random points in our training set. For each point we sampled a random radius between 500m and 1500m to create 1000 random regions. For each region we calculate the corresponding ratio of detections to visible firms; we obtain $mean(R_r)=0.91$1 and  $std(R_r)=0.04$. Therefore, although the detector might under-count the number of visible firms, there is no evidence of a distortion of the underlying spatial distribution.  

\section{Clustering}\label{clustersReg}

\begin{figure*}[t!]
\includegraphics[width=\linewidth]{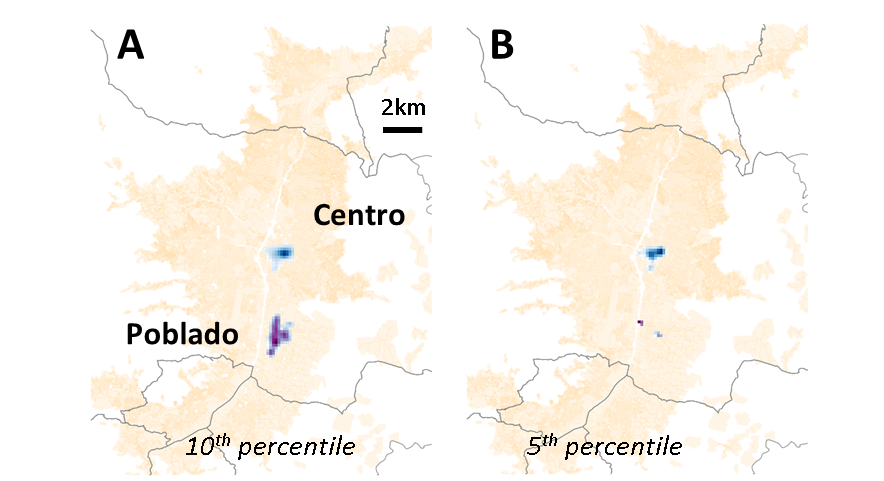}
\caption{Clusters of commercial registered firms. \textbf{(A)} When we take the top 10 percentile as a density threshold we identify two clusters: the city centre and the new business district in El Poblado. \textbf{(B)} If we consider the top 5 percentile, only the city centre remains as a well defined cluster. \label{regclust}}
\end{figure*}

Using the same methodology as in the main text \cite{fotheringham2000quantitative, arribas2014validity, anselin1995local}, we locate the clusters of commercial registered firms. Figure \ref{regclust} illustrates our findings: if we consider a density threshold of the top 10 percentile, we clearly see two clusters, the city centre and El Poblado; if we take the top 5 percentile instead we still clearly see the city centre, but El Poblado is only sparsely represented. These results hold whether we use the set of registered firms or the set of commercial registered firms.

\begin{figure*}[t!]
\includegraphics[width=\linewidth]{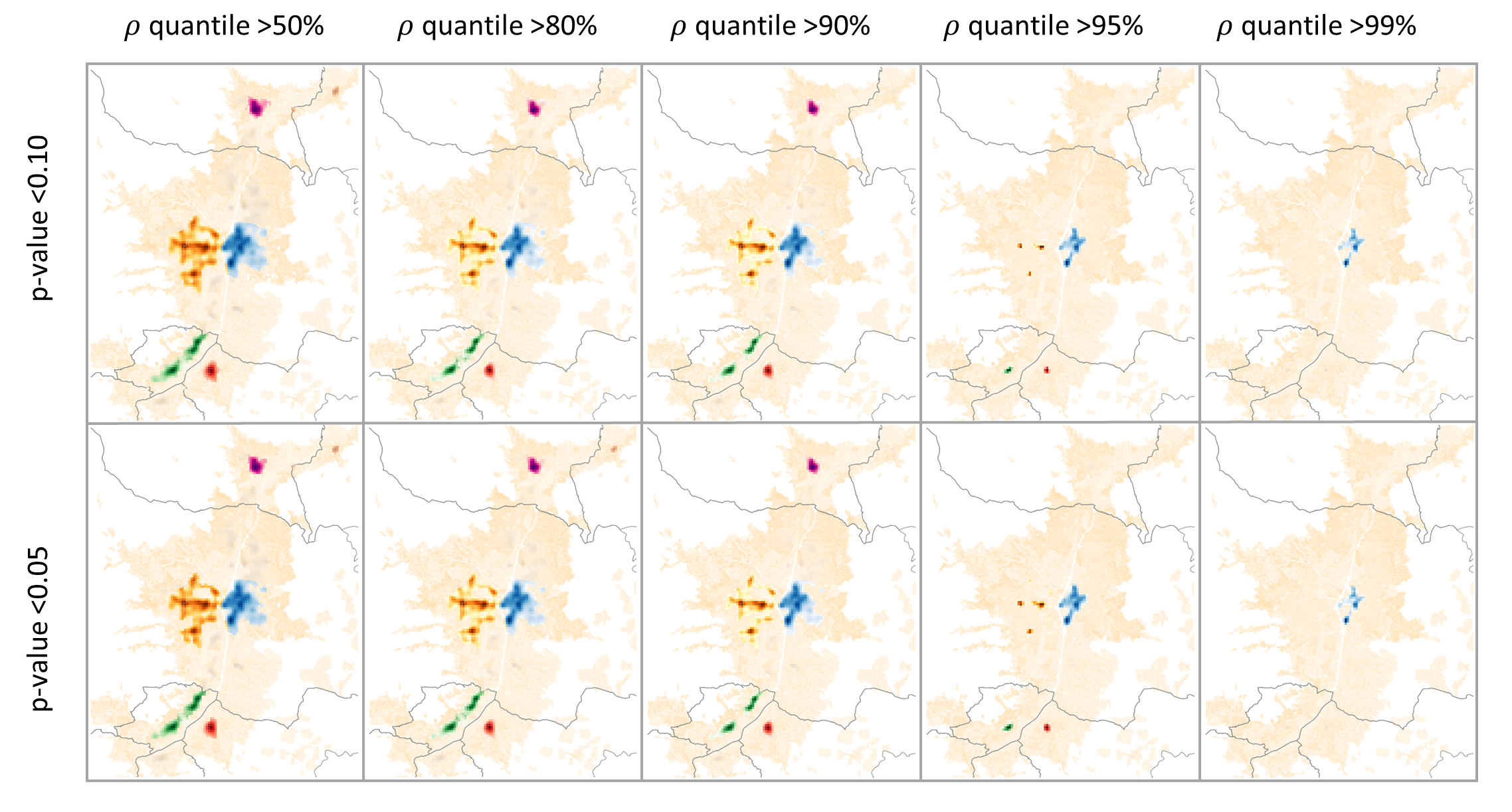}
\caption{We build a matrix of possible clusters, where each row represents a choice of p-value for the auto-correlation, and each column represents a different choice of density threshold. It can be seen that changing the p-value from 0.10 to 0.05 has little effect on the clusters identified for any given density threshold. Furthermore, there is no change in cluster composition when reducing the density threshold from the $80^{th}$ to the $50^{th}$ percentile.}
\label{robustclusters}
\end{figure*}

In the main text, we used a p-value of 0.10 of the local indicator of spatial auto-correlation to identify clusters of visible firms. At the same time, we varied the density threshold between the 80th and the 99th percentile to evaluate the persistence of clusters. 

Here we show that our results are robust with respect to the choice of p-value, and that relaxing the density threshold below the 80th percentile has little effect on the resulting clusters. To do so, we replicate our analysis using a p-value of 0.05. Furthermore, we also consider a density threshold at the median, which implies that half of the cells are labelled as showing high density of commercial activity. 

Figure \ref{robustclusters} shows that there is no difference between the number of clusters identified with a p-value of 0.05 and those identified with a p-value of 0.10. Furthermore, lowering the density threshold below 80\% has little effect on the cluster composition, besides a small increase in the area they cover.

\section{The commercial sector}\label{byindustry}

\begin{figure*}[ht!]
\includegraphics[width=\linewidth]{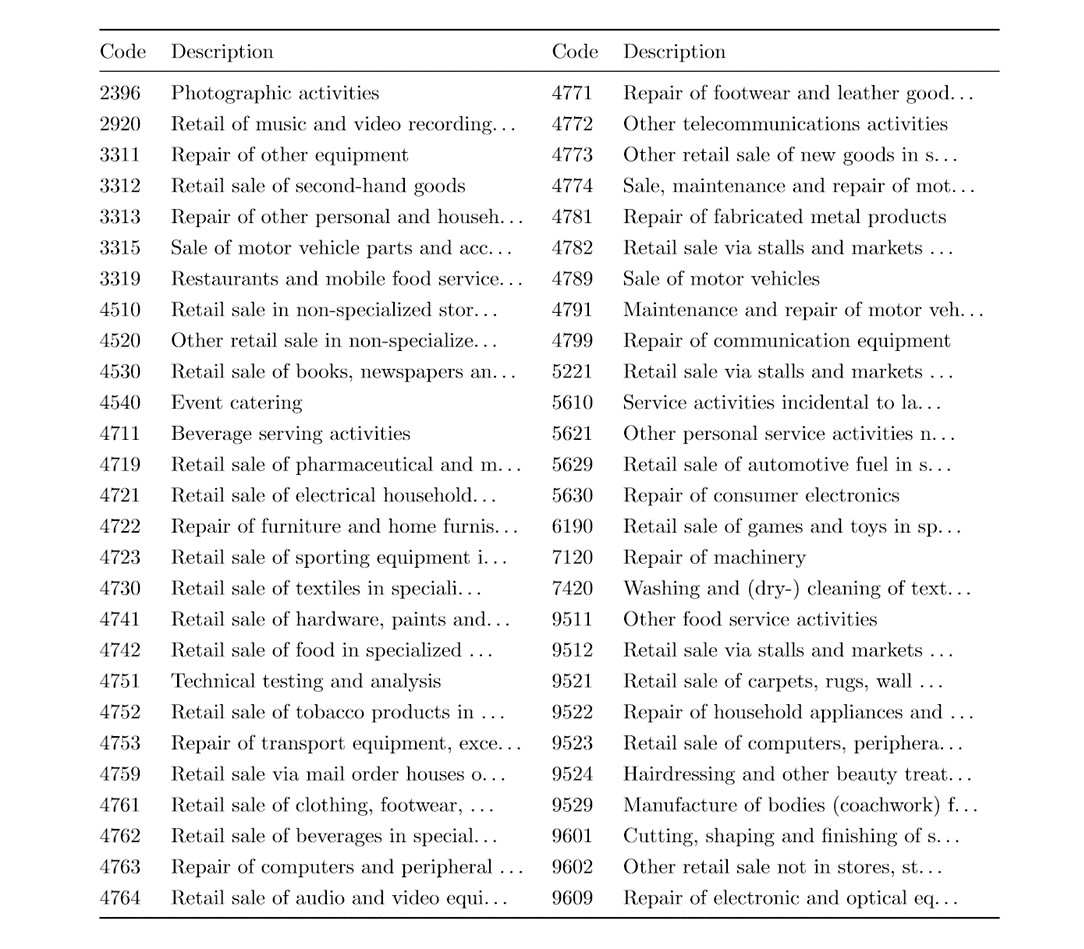}
\caption{List of all the industries (at the four digit level) which we use to identify registered commercial activity.}
\label{byindustryF}
\end{figure*}

We have used the term commercial registered firms to refer to firms which belong to industries that are likely to have shopfronts. The industries we used are based on the definition of street commerce found in \cite{sevtsuk2020street}. Table \ref{byindustryF} shows the list of these industries (at the four digit level). 

\section{Diversity}\label{diver}

\begin{figure*}[h!]
\includegraphics[width=\linewidth]{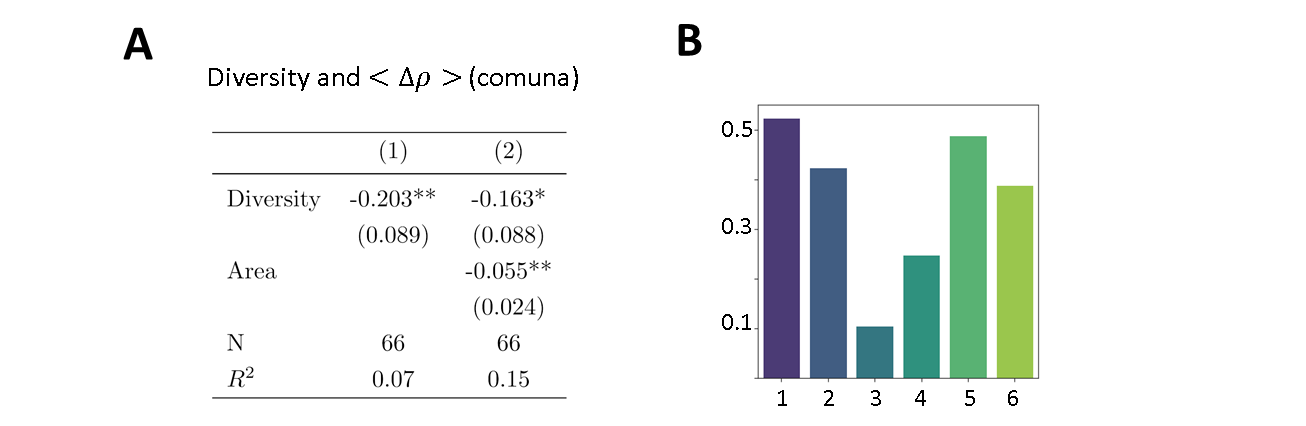}
\caption{\textbf{(A)} Summary of the regression of industrial diversity against $\Delta\rho$ at the comuna level. \textbf{(B)} Non-adherence of registered firms. We see the same patterns as for commercial registered firms.}
\label{divAdh}
\end{figure*}

In figure \ref{complexity} we showed that there is a significant relationship between the industrial diversity and $\Delta\rho$ at the neighbourhood level. We replicate the analysis, but this time we look at the comuna level. As in the previous case, the relationship is significant with a negative sign. The regression is summarized in Table \ref{divAdh}A. 

\section{Relationship between $\Delta\rho$ and socio-economic variables }\label{unweighted}

\begin{figure*}[h!]
\includegraphics[width=\linewidth]{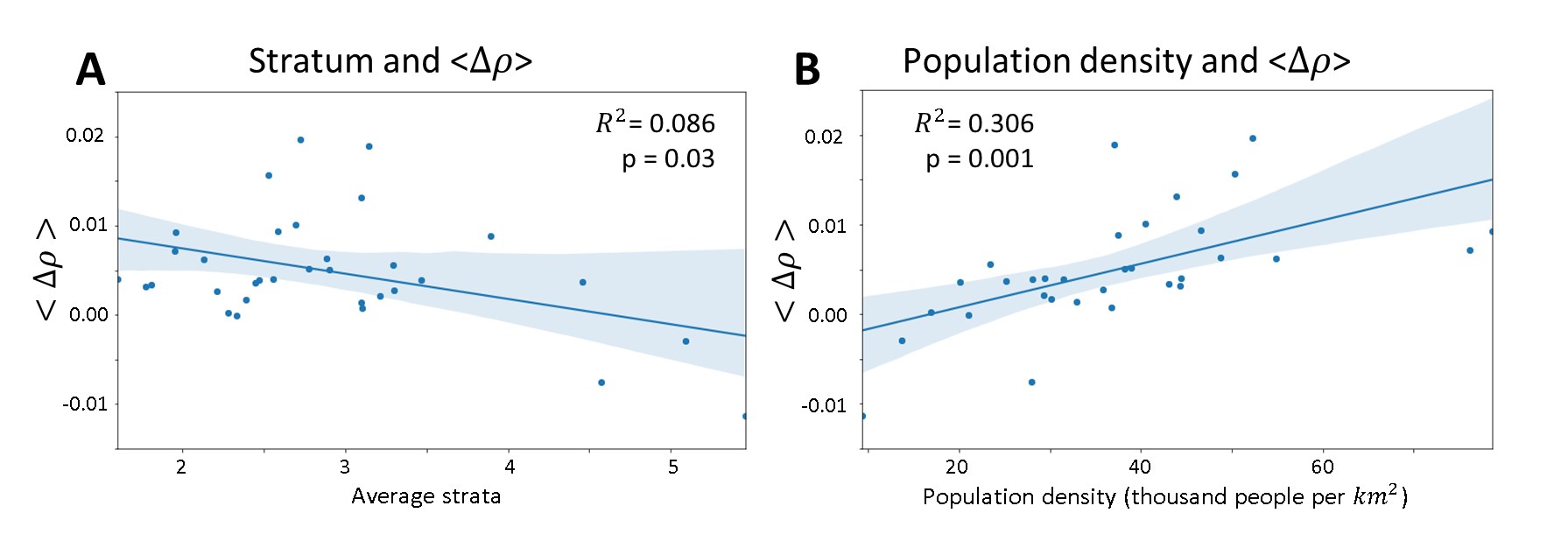}
\caption{$<\Delta\rho>$ and socio-economic characteristics. \textbf{(A)} We show the result of running an unweighted regression of $<\Delta\rho>$ against the average stratum at the comuna level. \textbf{(B)} Similarly, we show the result of running an unweighted regression of $<\Delta\rho>$ against the population density at the comuna level. We see that both relationships are significant at the $0.05$ level.}
\label{unwreg}
\end{figure*}

We had previously shown that $<\Delta\rho>$ has a strong significant relationship with both the stratum and the population density of the comuna. We replicate the exercise but using ordinary least squares instead of weighted least squares. The results can be seen in figure \ref{unwreg}C. In both cases, the regressions remain significant at the $0.05$ level, although the weighted regressions show a better fit.

\section{Land use of registered firms}\label{landuseReg}

We replicate the analysis shown in figure \ref{land}C, but using the full  registered firms set instead, to calculate the level of non-adherence of registered firms. In the main text, we showed that both visible firms and commercial firms show a high level of non-adherence to land use plans across all strata. The situation is similar for registered firms in general. Figure \ref{divAdh}B shows the fraction of registered firms that are located outside mixed land use. While the levels of non-adherence are lower than those of visible firms, they remain high, and there is little difference between the poor and the rich strata, although, as was the case for commercial registered firms, the two middle strata show a much higher adherence.

\end{document}